# *In situ* quasi-elastic neutron scattering study on the water dynamics and reaction mechanisms in alkali-activated slags

Kai Gong[a,b], Yongqiang Cheng[c], Luke L. Daemen[c], Claire E. White[a,b*]

[a] Department of Civil and Environmental Engineering, Princeton University, Princeton NJ 08544, USA

[b] Andlinger Center for Energy and the Environment, Princeton University, Princeton NJ 08544, USA

[c] Chemical and Engineering Materials Division, Oak Ridge National Laboratory, Oak Ridge TN 37830, USA

* Corresponding author: Phone: +1 609 258 6263, Fax: +1 609 258 2799, Email: whitece@princeton.edu

Postal address: Department of Civil and Environmental Engineering, Princeton University, Princeton NJ 08544, USA

## 1  Abstract

In this study, *in situ* quasi-elastic neutron scattering (QENS) has been employed to probe the water dynamics and reaction mechanisms occurring during the formation of NaOH- and $Na_2SiO_3$-activated slags, an important class of low-$CO_2$ cements, in conjunction with isothermal conduction calorimetry (ICC), Fourier transform infrared spectroscopy (FTIR) analysis and $N_2$ sorption measurements. We show that the single ICC reaction peak in the NaOH-activated slag is accompanied with a transformation of free water to bound water (from QENS analysis), which directly signals formation of a sodium-containing aluminum-substituted calcium-silicate-hydrate (C-(N)-A-S-H) gel, as confirmed by FTIR. In contrast, the $Na_2SiO_3$-activated slag sample exhibits two distinct reaction peaks in the ICC data, where the first reaction peak is associated with conversion of constrained water to bound and free water, and the second peak is accompanied with conversion of free water to bound and constrained water (from QENS analysis). The second conversion is attributed to formation of the main reaction product (i.e., C-(N)-A-S-H gel) as confirmed by FTIR and $N_2$ sorption data. Analysis of the QENS, FTIR and $N_2$ sorption data together with thermodynamic information from the literature explicitly shows that the first reaction peak is associated with the formation of an initial gel (similar to C-(N)-A-S-H gel) that is governed by the $Na^+$ ions and silicate species in $Na_2SiO_3$ solution and the dissolved Ca/Al species from slag. Hence, this study exemplifies the power of *in situ* QENS, when combined with laboratory-based characterization techniques, in elucidating the water dynamics and associated chemical mechanisms occurring in complex materials, and has provided important mechanistic insight on the early-age reactions occurring during formation of two alkali-activated slags.

## 2 Introduction

Alkali-activated materials (AAMs) are a class of sustainable cements synthesized by mixing aluminosilicate precursors with alkaline activating solutions, where the precursor particles dissolve and reprecipitate to form an interconnected gel network. When properly formulated, the resulting AAM binders exhibit favorable mechanical properties similar to hydrated ordinary Portland cement (OPC),[1] that is the main binder material used in concrete production. Due to the massive usage of OPC around the world (~4.1 billion tons in 2017),[2] the cement industry alone is responsible for approximately 5-9% of the global anthropogenic $CO_2$ emissions.[3-5] The negative sustainability aspects of OPC have catalyzed the research and development of alternative cementitious binders, with AAMs being one of the most promising candidates.[3] Compared with OPC, AAMs can have a substantially lower $CO_2$ emissions (up to ~80%[4]) as they are produced using precursor powders from industrial by-products (e.g., blast furnace slag and coal-derived fly ash) and naturally abundant ashes and clays.[5,6] In addition to the sustainability benefits outlined above, certain AAMs exhibit superior thermal properties and higher resistance to chemical attack (e.g., sulfate attack) when compared with OPC-based materials.[5,7,8]

The formation process of AAMs consists of dissolution of precursor particles and precipitation of reaction products, where dissolution and precipitation reactions occur concurrently, similar to the OPC hydration process. This formation process not only controls the early-age properties of AAMs (e.g., workability, setting, and hardening) and subsequent strength development but also dictates evolution of the pore structure and associated long-term durability performance. Many studies have shown that the formation processes of AAMs, along with their early-age and long-term properties, vary considerably depending on the precursor attributes (chemical and physical properties), activator chemistry and curing conditions (e.g., ambient or elevated temperature)[7, 9-15]. However, the exact formation mechanisms occurring during the alkali-activation reaction involving different precursor attributes and activator chemistries remain somewhat unknown, although recent studies have provided important mechanistic insight on specific AAM systems.[12, 16-21] One challenge that limits our ability to elucidate the exact formation mechanisms occurring on the order of hours (as opposed to days) is the need for experimental tools that allow for these mechanisms to be probed *in situ* and in a non-destructive manner.

The most common characterization technique used to study *in situ* reaction kinetics of OPC and AAMs is isothermal conduction calorimetry (ICC), which measures heat flow during the hydration or activation process.[10-12, 20, 22] This technique has greatly improved our understanding of OPC reaction kinetics, and is ideal for pin-pointing when the main binder phase, calcium-silicate-hydrate (C-S-H) gel, precipitates during hydration of OPC.[22] However, for formation processes that possess multiple reaction peaks in the heat flow data, such as an AAM based on silicate activation of blast furnace slag,[9, 13, 15] ICC cannot determine the reaction type (i.e., dissolution versus precipitation) or reveal which reactants are involved. Other experimental characterization techniques that are capable of providing valuable insight on the reactions mechanisms in cement-based systems include neutron and X-ray scattering,[12, 23-39] pair distribution function (PDF) analysis,[17, 18, 34, 39, 40] X-ray nanotomography,[41] nuclear magnetic resonance (NMR),[42, 43] differential scanning calorimetry,[44] and Fourier transform infrared spectroscopy (FTIR).[21, 45] One key neutron scattering technique that has been widely used to study reaction kinetics and associated mechanisms in OPC-based materials is quasi-elastic neutron scattering (QENS),[23-27, 29, 37, 44] which traces the evolution of different $H/H_2O$ components (e.g., bound and free water) during the hydration process.

Owing to the extremely large incoherent neutron scattering cross section of hydrogen atoms (H-atoms) compared with the other elements in cementitious materials (e.g., Ca, Si, Al, Mg, O, Na, S, Fe), over 99% of the signal in a QENS measurement is attributed to the dynamic behavior of H-atoms in these systems.[23] Furthermore, chemically bound or constrained H-atoms contribute to the QENS spectrum differently from that of mobile H-atoms, and, as a result a QENS measurement allows for quantification of different $H/H_2O$ environments within a cementitious system. By tracing the evolution of these different $H/H_2O$ environments as a function of reaction time, *in situ* QENS provides a direct measure of the hydration reaction in a cementitious system where free water (that starts off as the mixing water) is converted to (i) chemically bound $H/H_2O$ associated with the reaction products (e.g., interlayer $H_2O$ in C-S-H gel and OH units in both C-S-H and $Ca(OH)_2$ for OPC-based systems) and (ii) constrained water in gel pores or on pore surface.[23, 27, 31] Some of the original free water remains as such during the formation reaction, since large pores (i.e., capillary pores) containing $H_2O$ (and ions) also emerge as the reaction proceeds.[32, 46] *In situ* QENS has been shown to be particularly valuable for studying reaction kinetics and mechanisms when combined with simple nucleation/growth models and/or ICC measurements.[23-26, 29, 37] Another neutron scattering technique that is sometimes employed together with QENS is inelastic neutron scattering (INS),[28, 29] which is capable of probing the intramolecular vibrational and librational modes of $H_2O$ molecules [47] present in materials and also has been used for OPC-based materials to provide complementary data on $Ca(OH)_2$ formation.[28, 29] The use of *in situ* QENS to study low-$CO_2$ cementitious binders, such as AAMs, has so far been limited. To date, there is only one QENS study in the AAMs literature where *in situ* QENS was used to probe the formation process of a low-Ca fly ash-based AAM.[33]

In this study, *in situ* QENS and INS are employed for the first time to investigate the reaction kinetics and formation mechanisms of alkali-activated slags (AASs), a type of AAMs. We have collected *in situ* QENS and INS data for two AAS samples activated with different activator solutions (i.e., NaOH and $Na_2SiO_3$) during their initial ~8-12 hours of reaction. The *in situ* QENS and INS data have been fitted with several commonly used models, allowing for accurate quantification of the different $H/H_2O$ environments (chemically bound, constrained and free) present in the AASs as a function of reaction time. The resulting QENS water indices have then been compared with ICC, FTIR and $N_2$ sorption data to enable the dominant formation mechanisms to be elucidated. Hence, this investigation exemplifies the power of a *in situ* QENS for quantifying the evolution of different $H/H_2O$ environments in a complex chemical reaction occurring on the order of minutes to hours to days, shedding light on the dominant reaction processes occurring at different stages during the *in situ* measurement.

## 3 Materials & Methods

### 3.1 Materials

Two types of alkaline solutions (i.e., NaOH and $Na_2SiO_3$) were used to activate ground granulated blast furnace slag (hereinafter referred to as slag) with a chemical composition of ~33.9 wt. % CaO, ~37.0 wt. % $SiO_2$, ~9.0 wt. % $Al_2O_3$, and ~14.3 wt. % MgO.[48] The NaOH solution was prepared by dissolving NaOH pellets (Sigma-Aldrich, reagent grade) in distilled water (18 MΩ·cm), whereas the $Na_2SiO_3$ solution was prepared by adding NaOH pellets and distilled water to a commercial sodium silicate solution (Type D from PQ Corporation) to obtain a $Na_2O/SiO_2$ molar ratio of 1. The pastes were formulated to give a mix proportion of 7 g $Na_2O$ and 40 g water for 100 g of slag.

### 3.2 Experimental Details

*3.2.1 Quasi-elastic neutron scattering and inelastic neutron scattering (QENS-INS)*

Both the QENS and INS (denoted as QENS-INS) measurements were performed simultaneously on the VISION spectrometer at the Spallation Neutron Source (SNS), Oak Ridge National Laboratory (ORNL). The AAS sample was hand mixed for about 2 mins, and immediately after mixing, 5-6 g of the mixture was loaded into a cylindrical vanadium sample container. QENS-INS spectra were collected from ~−2 to ~1000 meV on the AAS mixture for ~8-12 hours at room temperature (i.e., ~300 K). The data have been binned every 5 min during the initial hour and every hour afterwards. It is estimated that the first data point was collected at approximately 10 min after mixing. A number of spectra were also collected on bulk water at different temperatures (i.e., 280-320 K) and supercooled ice at ~5 K. All the analysis was based on data from the backscattering bank, which has a $Q$-value of ~2.5 Å$^{-1}$ at the elastic line. This corresponds to a spatial resolution of $d$ = ~2.5 Å ($d = 2\pi/Q$), meaning that any diffusive motions within a ~2.5×2.5×2.5 Å$^3$ cage appear static to the instrument in this study. This Q-value is similar to what was used in several previous QENS studies on OPC-based systems (1.9-2.4 Å$^{-1}$).[23, 24, 28, 49]

*3.2.2 Isothermal conduction calorimetry (ICC)*

ICC measurements on the NaOH- and $Na_2SiO_3$-activated slags were conducted using a TAM Air isothermal calorimeter (TA Instruments). The AAS samples were prepared following the same mixing protocol adopted for the QENS-INS experiments. Immediately after mixing, ~5 g of AAS sample was transferred to a standard plastic container and then loaded in the calorimeter, together with a reference container with ~5 g of deionized water. Data were collected continuously for 24 hours at ~25 °C, and the heat released due to alkali-activation reaction was calculated by subtracting out the data from the reference container. An additional ICC test was performed on a slag-water mixture that had the same amount of slag and water as the two AAS samples, following the same mixing protocol outlined above.

*3.2.3 Fourier transform infrared spectroscopy (FTIR)*

Attenuated total reflectance (ATR)-FTIR measurements were performed on the AAS samples at different activation times over a period of 2-4 days after initial mixing, using a PerkinElmer FTIR instrument (Frontier MIR with a Frontier UATR diamond/ZnSe attachment) purged with an $N_2$ flow. For each measurement, 32 scans were collected from 4000 to 500 cm$^{-1}$ with a resolution of 4 cm$^{-1}$. To facilitate the assignment of the FTIR peaks in the AAS samples, FTIR spectra were also collected on the neat slag and activator solutions (i.e., NaOH and $Na_2SiO_3$).

*3.2.4 $N_2$ sorption*

$N_2$ sorption experiments were performed on the AAS samples that had reacted for ~6 and ~12 hours in order to examine the development of the pore structure during the early stages of reaction. After allowing the samples to undergo the alkali-activation reaction for the designated time in a sealed container, the samples were crushed and sieved to obtain particles with diameters of 0.5-1 mm. The particles were then soaked in an isopropyl alcohol bath (~1 g of sample per ~200 mL of isopropyl alcohol) for ~24 hours to stop the activation reaction and lower the surface tension during subsequent drying. Once filtered from the isopropyl alcohol, the particles were vacuum dried at room temperature for ~3-4 days using a vacuum pump. $N_2$ sorption measurements were performed on the samples using a Micromeritics 3Flex instrument after degassing the sample in the degassing port of the instrument for approximately 1 day at ~333 K.

## 4 Results & Discussion

### 4.1 Typical QENS-INS Spectra

Figure 1 shows a typical QENS-INS spectrum (from ~ −2 to 400 meV) collected on a freshly mixed $Na_2SiO_3$-activated slag sample using VISION (first data point collected at ~10 min after mixing), together with the QENS-INS spectra of bulk water at 290 K and ice Ih at 5 K from the same instrument. It is evident that the QENS-INS spectrum of the $Na_2SiO_3$-activated slag sample resembles that of the bulk water and is relatively featureless compared with the ice Ih spectrum, where well-defined features are clearly observed (e.g., hydrogen bond stretching and intermolecular translation modes, as labeled in Figure 1). These well-defined features in the ice Ih spectrum are attributed to the ordered structure of the ice Ih crystal and are consistent with data reported in the literature.[50, 51] In contrast, these well-defined features are absent in the $Na_2SiO_3$-activated slag sample, where the only apparent features in the spectrum are a sharp elastic peak centered at ~0 meV and a broad diffuse water librational peak located at ~40-140 meV, similar to the case of bulk water. This sharp peak at ~0 meV is a superimposition of elastic and quasi-elastic scattering response of neutrons, which is generally referred as quasi-elastic neutron scattering (QENS) in the literature. The elastic response of neutrons results from collisions with tightly bound $H/H_2O$ with motions slower than that of the instrument resolution, whereas the quasi-elastic response is mainly attributed to collisions with more mobile $H/H_2O$ in the sample. The former results in sharping of QENS while the latter leads to broadening of QENS, with higher mobility $H/H_2O$ giving a larger extent of quasi-elastic broadening.

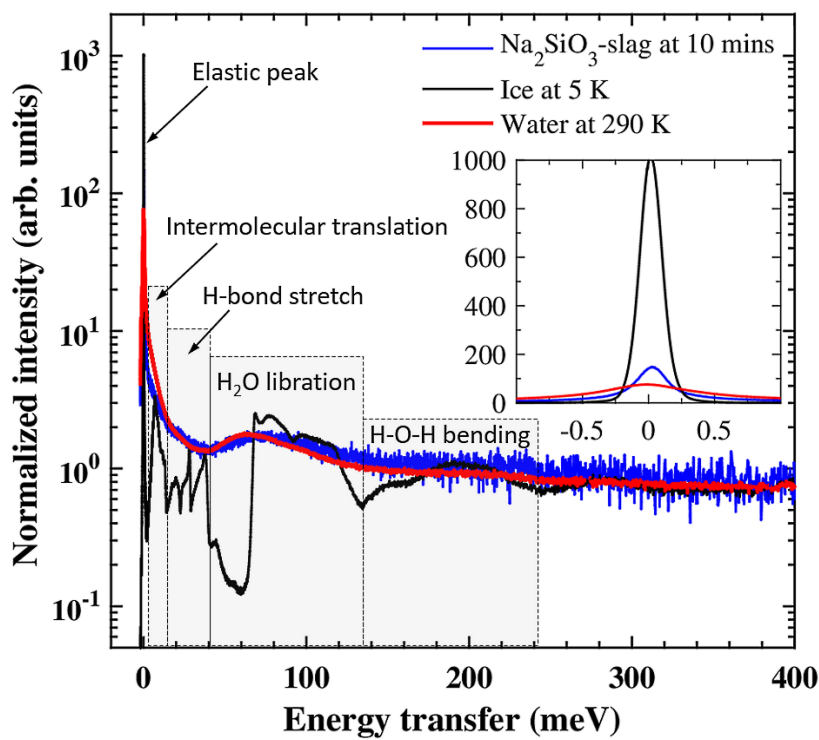

Figure 1. Comparison of the QENS-INS spectrum of a freshly mixed $Na_2SiO_3$-activated slag sample (data collected at approximately 10 min after mixing) with pure liquid water at 290 K and ice Ih at 5 K. To facilitate comparison, each data set has been normalized by the total area below the corresponding spectrum, which reflects the total amount of H-atoms in each sample. Note that the normalized intensity in the main figure is given on a logarithmic axis. The inset figure is a zoom of the elastic peak centered at ~0 meV on a linear axis.

A closer examination of the QENS component (inset in Figure 1) reveals that both the intensity and the extent of broadening for the $Na_2SiO_3$-activated slag sample lie between that of the 5 K ice Ih and the 290 K bulk water, where ice Ih exhibits a mainly elastic scattering response with little quasi-elastic broadening whereas the bulk water sample is dominated by quasi-elastic broadening. This behavior is expected since for 5 K ice Ih the H-atoms have a residence time much longer than what VISION can resolve (i.e., the H-atoms appear stationary to the instrument) and therefore these H-atoms scatter neutrons elastically. On the other hand, bulk water has a residence time of ~1.1-1.65 ps,[52-55] and hence the $H_2O$ molecules are mobile under the energy resolution of the current experiment (approximately 6.6 ps, as discussed in the next section), giving rise to the quasi-elastic broadening seen in Figure 1. The QENS component of the freshly mixed $Na_2SiO_3$-activated slag sample is narrower compared with that of bulk water, indicating that $H/H_2O$ in the former is less mobile. Previous studies have shown that a number of factors can slow down the mobility of $H/H_2O$, including lowering the temperature,[53, 54, 56, 57] ion solvation,[57, 58] interaction with solid surfaces,[52, 59-61] and confinement by nanoscale pores.[54, 56, 59, 62] Given that the $Na_2SiO_3$-activated slag sample was measured at a higher temperature than the bulk water sample (i.e., 300 K versus 290 K), the lower $H/H_2O$ mobility in the $Na_2SiO_3$-activated slag can be attributed to the presence of (i) ions (i.e., $Na^+$ and silicate species in the original $Na_2SiO_3$ solution along with any dissolved species from slag particle surface upon mixing) and (ii) solid surfaces (i.e., slag particles). A detailed analysis of the impact of these different factors on $H/H_2O$ mobility in freshly mixed AAMs will be presented in Section 4.3, based on fitting of the QENS spectra (presented in Section 4.2) and literature data.

In addition to the QENS component, the QENS-INS spectrum of the $Na_2SiO_3$-activated slag sample also clearly exhibits a visible water librational feature at ~40-140 meV (Figure 1), although this feature is not as well defined as the corresponding feature in the 5 K ice Ih sample. The position of the librational peak is an indication of rotational mobility of $H_2O$, with a higher energy transfer value associated with a lower rotational mobility.[47, 63] It is clear from Figure 1 that the librational feature of $Na_2SiO_3$-activated slag mixture is slightly shifted to a higher energy transfer value (peaked at ~72 meV) compared with bulk water (peaked at ~64 meV), indicating that the rotational mobility of $H_2O$ in the former is lower. This result agrees with the QENS data (as already discussed above) and is also consistent with previous INS studies where it has been shown that ion solvation and interaction with solid surfaces result in a higher energy transfer for the water librational peak position.[47, 50, 63]

### 4.2 Notes on Data Analysis

Figure 2 shows the evolution of the QENS component of the QENS-INS spectrum for the $Na_2SiO_3$-activated slag sample during the initial ~7.5 hours of reaction. It is seen that the QENS spectrum sharpens with the progress of reaction, due to a transition from mobile $H_2O$ (quasi-elastic response) to less mobile $H/H_2O$ (elastic response). Similar sharpening behavior of QENS spectra during the early stages of OPC hydration has been reported,[23, 30, 31] and is generally attributed to a conversion of mobile $H_2O$ to tightly bound/constrained $H/H_2O$. The data in Figure 2 illustrate that the QENS component of the QENS-INS spectrum collected at VISION contains a wealth of useful information that can be exploited with proper data analysis. Hence, although VISION is not a dedicated QENS instrument, the wide energy transfer coverage allows for QENS analysis to be carried out, in conjunction with examination of the higher energy $H/H_2O$ INS dynamics (e.g., water libration).

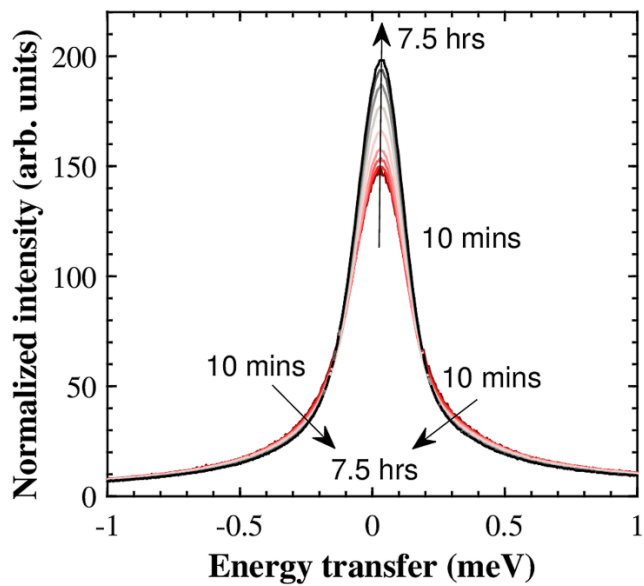

Figure 2. Evolution of the elastic peak for Na$_2$SiO$_3$-activated slag with the progress of reaction. These data have been normalized by the total area under the 1-hour QENS-INS spectrum of this sample (from ~ −2 to 1000 meV).

To quantify the changes occurring in the QENS-INS data, fitting of the QENS and water librational components of the QENS-INS spectra has been performed with commonly used models documented in the literature. A "Gaussian + Lorentzian" model, which has been extensively used in the past for QENS analysis of cementitious materials,[23-28, 49] has been employed here to fit this component of the spectrum. A double-Lorentzian model was chosen here mainly because a previous QENS study (with similar energy resolution and $Q$-value as the current study) on the hydration of pure tricalcium silicate (the major component of OPC) has explicitly showed that a double-Lorentzian model is more physically realistic than a single-Lorentzian model, where the two Lorentzians have been attributed to constrained and free water respectively.[24] We have fitted our QENS data with single-, double- and triple-Lorentzian models, and the results for all the samples (Figure S1 in the Supporting Information) further justify the selection of the double-Lorentzian model for the current study.

The double-Lorentzian model for the QENS response can be written as follows[24]:

$$S_{inc}(\omega) = \left\{A\delta(\omega \approx 0) + B_1\left[\frac{\Gamma_1}{\pi(\Gamma_1^2+\omega^2)}\right] + B_2\left[\frac{\Gamma_2}{\pi(\Gamma_2^2+\omega^2)}\right]\right\} \otimes \left[\left(\frac{1}{\sigma\sqrt{2\pi}}\right)e^{(-\omega^2/2\sigma^2)}\right] + (C + D\omega)$$

----------------------------------------------------------------------------------------------------------------- (1)

where $S_{inc}(\omega)$ is the incoherent scattering intensity, and $\omega$ is the energy transfer of the scattered neutrons. $\Gamma_1$ and $\Gamma_2$ are the half-width at half-maximum (HWHM) of the two Lorentzians. The instrument resolution is approximated by a Gaussian function, with $\sigma$ denoting the Gaussian standard deviation.[23, 24] The HWHM of the instrument resolution has been determined by fitting the QENS spectrum of 5 K ice Ih with a Gaussian, as illustrated in Figure 3a. It is seen that a Gaussian with HWHM of 0.10 meV is in excellent agreement with the 5 K ice Ih data, hence the instrument resolution has been fixed at this value for the remainder of this study. The energy resolution of 0.10 meV corresponds to a residence time of 6.6 ps, as determined using $\tau$ (ps) =

$\hbar/\Gamma$(HWHM).[55] The prefactors $A$, $B_1$ and $B_2$ dictate the intensity of the Gaussian and the two Lorentzians, respectively. $C + D\omega$ is a linear background, which is determined by fitting the 1-hour spectrum of each sample type and subsequently keeping the background fixed for the data fitting of other spectra.

The broader Lorentzian ($\Gamma_2$) in the above model represents free water, where the molecular motions of the water molecules are not hindered by any external constraints (e.g., ion solvation, interaction with solid surfaces and nanoconfinement). Hence, $\Gamma_2$ needs to be fixed to an experimentally determined value for a given temperature and $Q$-value, as has been done in a previous QENS study on OPC-based systems.[24] To determine $\Gamma_2$, we have analyzed several bulk water data sets that were collected at different temperatures on the VISION instrument by fitting the QENS component using a single Lorentzian convoluted with the instrument resolution function (approximated by the Gaussian shown in Figure 3a) together with a linear background. The quality of one fit is illustrated in Figure 3b, and the fit results are presented in Figure 3c, where it is clear that the HWHM of bulk water increases almost linearly with the measurement temperature, and a temperature of 300 K gives a HWHM of ~0.54 meV. This value aligns with previous QENS measurements where HWHM values of 0.50-0.75 have been reported for bulk water at similar temperatures and $Q$-values.[24, 56, 64] Furthermore, a HWHM of 0.54 meV corresponds to a residence time of ~1.20 ps, which is also in general agreement with residence time measurements of bulk water from other experiments (~1.10-1.60 ps).[54, 55] Hence, the HWHM of the broader Lorentzian (i.e., $\Gamma_2$) is fixed at 0.54 meV during the fitting process of the AAS samples while four parameters in the double-Lorentzian model (i.e., $A$, $B_1$, $B_2$ and $\Gamma_1$) are refined.

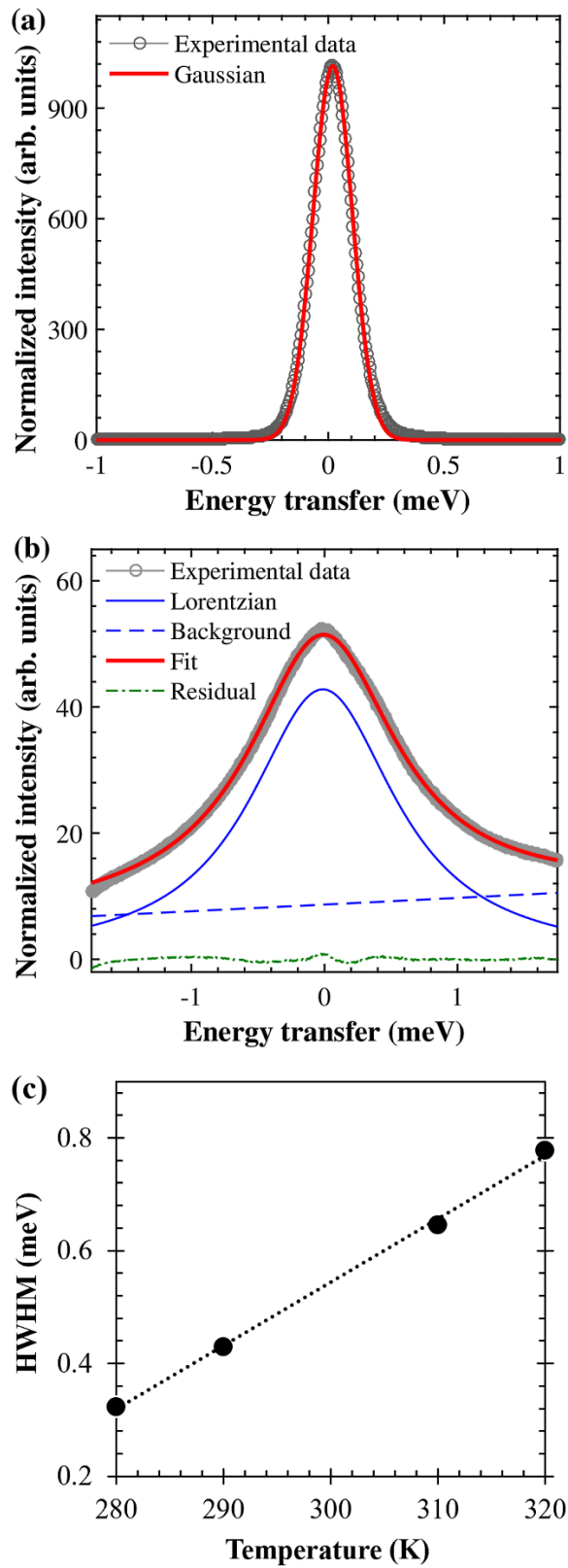

Figure 3. Typical QENS peak of (a) ice Ih at 5 K, and (b) bulk liquid water at 310 K from QENS-INS measurement on the VISION instrument. A Gaussian function has been used to fit the ice data

in (a), whereas a Lorentzian convoluted with the Gaussian determined in (a) was used to fit the bulk water data in (b). (c) shows the impact of temperature on the HWHM of the Lorentzian of bulk water, as determined using the model shown in (b).

A typical fit of the QENS data for the $Na_2SiO_3$-activated slag sample with the double-Lorentzian model is illustrated in Figure 4, which shows excellent agreement between the experimental data and the model. Based on the fit results, we have calculated the bound water index (BWI), constrained water index (CWI) and free water index (FWI), which are defined as follows:

$$BWI = \frac{A_G}{A_G + A_{L1} + A_{L2}}$$

$$CWI = \frac{A_{L1}}{A_G + A_{L1} + A_{L2}}$$

$$FWI = \frac{A_{L2}}{A_G + A_{L1} + A_{L2}}$$

where $A_G$, $A_{L1}$, $A_{L2}$ are the areas under the Gaussian and the two Lorentzians, respectively. Hence, the BWI accounts for the fraction of $H/H_2O$ in the system that have motions (rotation or diffusion) slower than the energy resolution of the instrument (i.e., 0.10 meV for HWHM of Gaussian as determined in Figure 3a). The FWI represents the fraction of $H_2O$ in the system that have motions similar to bulk free water, whereas CWI denotes the rest of $H/H_2O$ in the system that have motions faster than the instrument resolution but slower than bulk free water.

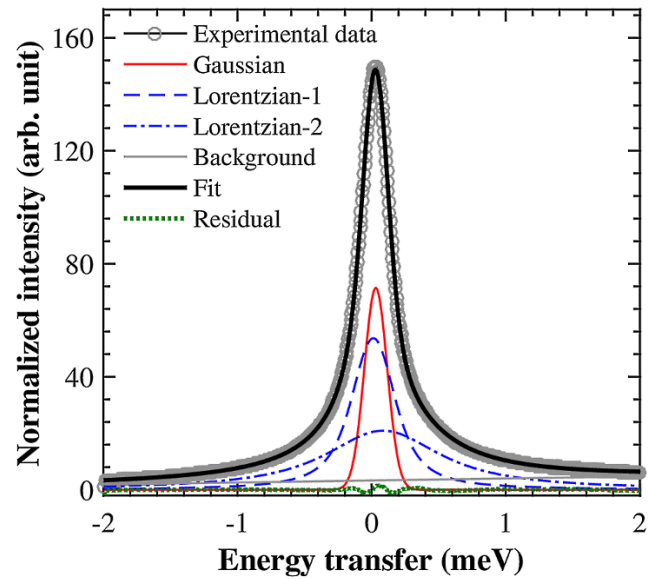

Figure 4. Typical fit of the QENS component for the $Na_2SiO_3$-activated slag sample (1 hour after mixing) with the double-Lorentzian model.

The broad librational peak in the INS spectrum is a reflection of rotational mobility of water molecules in the system, and this peak can be deconvoluted into three Gaussians (as shown in Figure 5), which account for the rock, wag, and twist modes of water libration, respectively.[63] The weighted librational peak position (WLPP) has been used to quantitatively compare the librational peak of each sample, which is defined as:

$$WLPP = \frac{A_{G1} \cdot P_{G1} + A_{G2} \cdot P_{G2} + A_{G3} \cdot P_{G3}}{A_{G1} + A_{G2} + A_{G3}}$$

where $A_{G1}$, $A_{G2}$ and $A_{G3}$ are the areas under the three Gaussians, and $P_{G1}$, $P_{G2}$ and $P_{G3}$ are the peak positions of the three Gaussians, respectively. The WLPP denotes the weighted average energy required to excite the water libration mode, with a higher WLPP value indicating lower water librational mobility. The evolution of WLPP with the progress of reaction for the two AASs, and its correlation with the QENS data are presented and discussed in the Supporting Information (Figure S2).

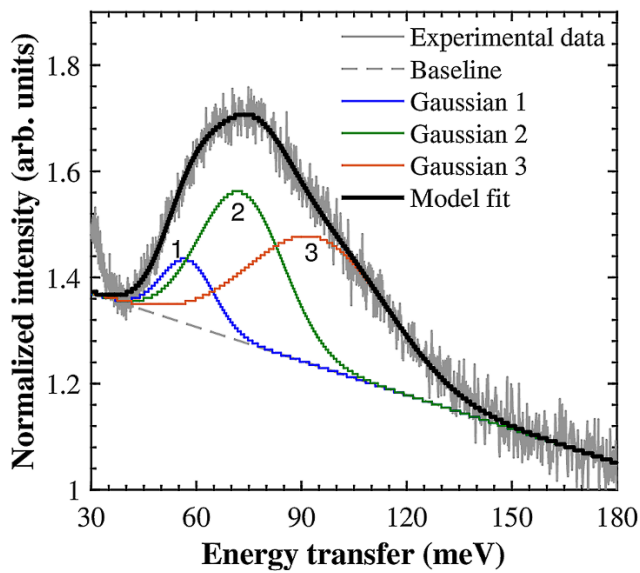

Figure 5. Deconvolution of the broad librational peak from INS for the $Na_2SiO_3$-activated slag sample at 1 hour after mixing.

## 4.3 Water Dynamics in Alkaline Activation Solutions

There are many different $H/H_2O$ environments present in AASs, as summarized in Table 1. It is important to understand how these $H/H_2O$ environments contribute to the different water indices (i.e., BWI, CWI, and FWI) introduced in Section 4.2, before examining the evolution of these water indices with the progress of the alkali-activation reaction. This understanding can be obtained by combining literature data on the residence time of the different $H/H_2O$ environments (as summarized in Table 1) with analysis of the QENS data for freshly mixed AAMs (~10 min after mixing), where the QENS data are dominated by the properties of the original activator solution due to the limited extent of reaction.

Table 1. Summary of the main types of $H/H_2O$ environments present in AASs and their assignments to different water indices.

| Major $H/H_2O$ environments in AASs | Residence time from literature | | Contribution to water indices | | |
|---|---|---|---|---|---|
| | Experiments | Molecular dynamics (MD) simulations | BWI (> ~6.6 ps) | CWI (< ~6.6 ps) | FWI (~1.3 ps) |
| $H/H_2O$ chemically bound in the reaction products (e.g., C-(N)-A-S-H) | > ~100 ps[55] | | Yes | - | - |
| H-atoms chemically bound in silicate monomer/oligomers | N/A | N/A | Yes | - | - |
| $H_2O$ adsorbed on solid surface of reaction products | | ~5-1000 ps[52, 65, 66] | Yes | Yes | - |
| $H_2O$ solvating $Na^+$ | | ~10-25 ps[67-70] | - | Yes | - |
| $H_2O$ solvating silicate monomer/oligomers | N/A | N/A | Yes | Yes | - |
| $H_2O$ confined in pores of ~2-3 nm | ~2-3 ps[64] | ~2-5 ps | - | Yes | - |
| Free $H_2O$ in large capillary pores | | | - | - | Yes |

As shown in Table 1, there are primarily three types of $H/H_2O$ in AASs that contribute to BWI, where the motions of $H/H_2O$ are slower than the instrument resolution. The first major contributor to BWI is chemically bound $H/H_2O$ that is incorporated into the structure of the reaction products. Due to the high Ca content in blast furnace slag, the main binder phase formed in AAS is a sodium-containing aluminum-substituted calcium-silicate-hydrate (C-(N)-A-S-H) gel, which contains a large portion of chemically bound H, similar to the C-S-H gel in OPC-based systems.[5, 24, 30] In addition to the C-(N)-A-S-H gel, secondary product phases (e.g., hydrotalcite-like and AFm phases) are often observed in AASs,[5, 9, 12, 48, 71] which also contain chemically bound H/$H_2O$. These chemically bound H/$H_2O$ in solid phases have a residence time longer than ~100 ps,[55] which is much longer than what the instrument can resolve (~6.6 ps).

The second contribution to the BWI are the H-atoms in silicate monomers/oligomers residing in the AAS pore solution. There is a lack of QENS data or other studies in the literature reporting the residence time of H-atoms in silicate monomers/oligomers. Nevertheless, analysis of the BWI for the initial activator solutions (Figure 6a) shows that the BWI increases significantly as the Si concentration of the activator solution increases. This clearly suggests that the H-atoms in silicate monomers/oligomers contribute to BWI. In addition, NMR studies on aqueous $Na_2SiO_3$ solutions showed that silicate monomers have a self-diffusion coefficient of ~$1.9 \times 10^{-10}$ $m^2/s$,[72] which is more than an order of magnitude lower than that of bulk water at ~296 K (~$2.3 \times 10^{-9}$ $m^2/s$).[73] It has been shown that the residence time of bulk water increases from ~1.1 to ~23 ps when its self-diffusion coefficient is lowered to ~$4.0 \times 10^{-10}$ $m^2/s$.[54] Hence, these findings suggest that the H-atoms in silicate monomers/oligomers have a residence time longer than ~23 ps, meaning that it is reasonable to assign these H-atoms to the BWI.

H-atoms in aluminate monomers and alumina-containing oligomers should also contribute to the BWI,[74] however, their contribution is likely to be relatively small compared with the H-atoms in silicate species because the aluminate concentration in the pore solution of AASs is at least 1-2 orders of magnitude smaller than the silicate concentration according to thermodynamic calculations.[71] Another contribution to the BWI in AASs is $H_2O$ adsorbed on the surface of the dissolving precursor particles and the precipitating reaction products (e.g., slag and C-(N)-A-S-H gel), since translational residence times of ~100-1000 ps have been reported for water molecules on C-S-H/silica surfaces.[52, 65, 66] However, one of these investigations (utilizing molecular dynamics (MD) simulations[65]) showed that the rotational residence time of $H_2O$ molecules on silica surfaces is only 2 times higher than that of bulk water, which suggests that the adsorbed $H_2O$ on solid surfaces may also contribute to the CWI.

Other types of H/$H_2O$ in the AAS that are assigned to CWI are mainly $H_2O$ molecules solvating ionic species in the pore solution (i.e., $Na^+$ and silicate species), as illustrated in Table 1. It is noted that there are also $Mg^{2+}$ and $Ca^{2+}$ ions and aluminate monomers in the AAS pore solution, however, the $H_2O$ molecules solvating these ions/species are not considered here because these ions/monomers are at much lower concentrations compared with $Na^+$ and silicate species in the AAS pore solution.[71] The impact of $Na^+$ on water dynamics has been examined using MD simulations which revealed that the residence time of $H_2O$ molecules in the 1st solvation shell of $Na^+$ is ~10-25 ps,[57, 68-70] suggesting that the $H_2O$ molecules solvating $Na^+$ ions should be assigned to the BWI. However, Figure 6a clearly shows that the BWI of the three NaOH solutions are similar despite their large difference in concentrations (~5.6, ~8.2 and ~10.0 M). In contrast, the CWI of these NaOH solutions is seen to increase almost linearly with increasing NaOH concentration (Figure 6b). These results show that the $H_2O$ molecules solvating $Na^+$ actually contribute to the CWI. The large residence time given by the MD simulations can be attributed to the fact that MD tends to overestimate the residence time of $H_2O$. For instance, many MD studies have shown that the residence time of bulk water at room temperature is about 3-10 ps,[57, 60, 65, 68, 69] which is about 2-5 times higher than those obtained experimentally (~1.1-1.6 ps).[54, 55] Therefore, it is reasonable to assign $H_2O$ molecules solvating $Na^+$ ions to the CWI based on the above observations in Figures 6a and 6b.

Figure 6b also shows that the 2.8 M $Na_2SiO_3$ solution has a much higher CWI than the 5.6 M NaOH solution, although their Na concentrations are the same (5.6 M). Hence, the $H_2O$ molecules solvating silicate monomers/oligomers in the 2.8 M $Na_2SiO_3$ solution should be assigned to CWI. Another observation in Figure 6b that supports this assignment is that the data point for the 2.8 M $Na_2SiO_3$ solution falls within the linear increasing trend observed for the three NaOH solutions and bulk water. However, at higher concentrations of $Na_2SiO_3$ (i.e., 4.1 and 5.0 M), where the Na+Si concentration is high (i.e., 12.3 and 15.0 M), the CWI is seen to decrease with Na+Si concentration (Figure 6b). This could be attributed to a reduction in the number of $H_2O$ molecules that are able to fully solvate each ion (refer to Supporting Information for calculations), leading to some of the $H_2O$ molecules to be transformed from CWI to BWI. This transformation also explains the exponential increase (as opposed to linear increase) of BWI with increasing concentration of Na$_2$SiO$_3$ as shown in Figure 6a. Furthermore, it is seen in Figure 6c that FWI is inversely correlated with Na+Si concentration of the solution, with a higher concentration leading to a smaller FWI.

Confinement of H$_2$O in nanosized pores also has a large impact on the H$_2$O dynamics, as has been shown in many previous QENS and INS studies.[54, 56, 59, 62] A recent QENS measurement on porous silica and zeolites showed that confinement in ~2-5 nm pores increases the residence time of ~296-300 K water from ~1.1 ps to ~2-4 ps (averaged over all the confined water molecules in the pores),[54, 64] which would contribute to the CWI in this investigation. During the initial period of AAS formation where the extent of reaction is small, the impact of confinement is negligible due to the absence of a well-developed pore structure. However, with the progress of reaction, small gel pores in the range of ~2-5 nm will start to develop as shown in a recent small-angle neutron scattering (SANS) study[32] as well as our N$_2$ sorption measurements on the same AAS samples that have reacted for 6 and 12 hours (see Section 4.5.2).

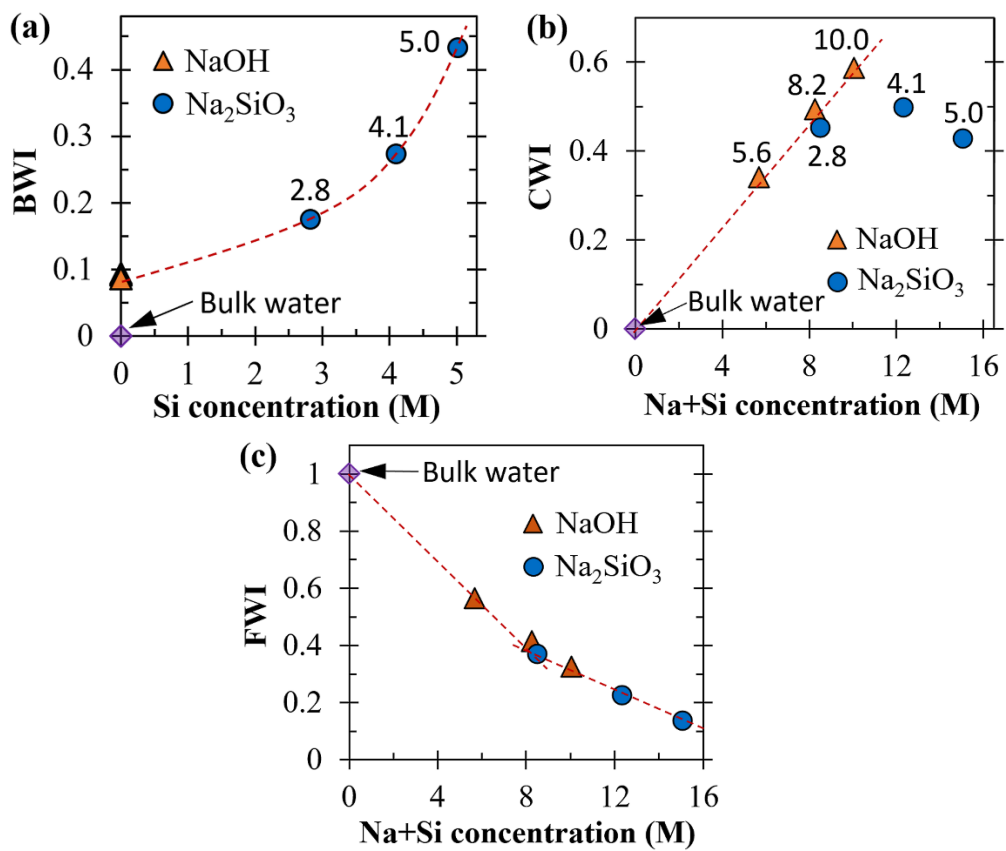

Figure 6. (a) Relationship between bound water index (BWI) of the freshly mixed alkali-activated materials (~10 min after mixing) and the Si concentration of the corresponding activator solutions. (b) and (c) show the correlations between constrained water index (CWI) and free water index

(FWI) of the same freshly mixed alkali-activated materials and the Na+Si concentration of the corresponding activator solutions, respectively. The indices for bulk water at 300 K are also given in the figures. The red dashed lines are provided to guide the eye.

## 4.4  Isothermal Conduction Calorimetry (ICC)

ICC is a commonly used technique to probe the kinetics of reaction for AAMs,[10, 11, 13-15, 75] and is employed here to provide complementary information for the QENS data. Figure 7 shows the ICC heat flow and cumulative heat curves during the initial ~12 hours of reaction for the NaOH- and $Na_2SiO_3$-activated slags. Both samples exhibit an intense heat flow peak immediately after mixing (within the first 10 min), which is commonly observed in both AAM and OPC-based systems.[12, 22, 75] This initial peak has generally been attributed to dissolution and wetting of particle surface upon mixing in cementitious systems.[12, 22] However, Figure 7b shows that this initial intense heat flow peak resembles that of the slag-water mixture where it is known that slag dissolution and precipitation reactions are much slower in water than in alkaline solutions.[6, 76] These results, along with the ICC data on a metakaolin-water mixture (refer to Figure S3 and the discussion in the Supporting Information), suggest that the initial heat flow in AASs is dominated by wetting of particle surfaces, however, a detailed analysis and discussion of this behavior is beyond the scope of the current paper. It is shown in Figure 7 that, in addition to the initial wetting peak, the NaOH-activated slag exhibits a single reaction peak (centered at ~3 hours, see Figure 7a) while the $Na_2SiO_3$-activated slag exhibits two reaction peaks (one at ~40 min and the other at ~6 hours, as visible in Figure 7a). The observation of single and double reaction peaks for the two types of AASs are consistent with previous ICC studies on hydroxide-[12, 15] and silicate-activated slags,[9, 13, 15] although their peak positions differ considerably depending on the curing temperature, activator chemistry (i.e., alkalinity), and slag chemistry.[9, 13, 15, 75] As reported in the literature, the single reaction peak for the NaOH-activated slag is attributed to precipitation of the main binder gel (i.e., C-(N)-A-S-H),[12] which is also the case for the second reaction (at ~6 hours) for the $Na_2SiO_3$-activated slag.[14] However, the exact cause of the first reaction peak in $Na_2SiO_3$-activated slag is unknown. Although several studies[13-15, 77] have postulated the formation of a "primary C-S-H" during the first reaction peak in this system, resulting from the reaction between Si species from the original $Na_2SiO_3$ solution and dissolved $Ca^{2+}$ from the slag surface, direct experimental evidence is needed in support of this claim.

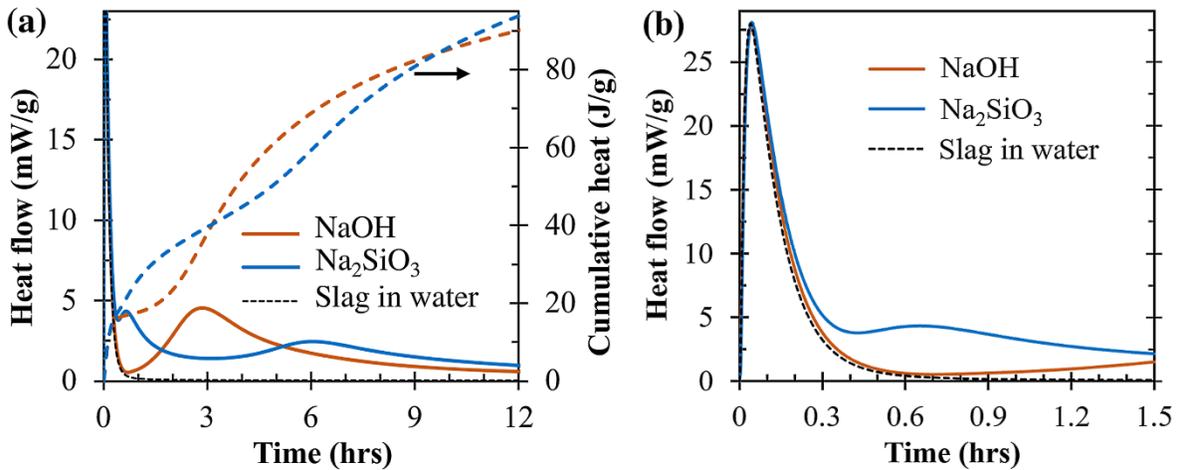

Figure 7. (a) Heat flow and cumulative heat for the NaOH- and Na$_2$SiO$_3$-activated slags during the initial 12 hours of reaction, obtained using ICC. The data for a slag-water mixture are also shown. (b) is a zoom of the initial 1.5 hours of reaction.

## 4.5 Evolution of QENS Water Indices

### 4.5.1 NaOH-activated slag

The evolution of BWI with the progress of reaction for the NaOH-activated slag is presented in Figure 8a, which shows that the BWI increases continuously with time, with a higher rate of increase during the initial ~4.5 hours and lower rates thereafter. This trend of evolution follows closely with the cumulative heat curve from ICC as also shown in Figure 8a, clearly demonstrating that the single reaction peak in this sample (Figure 7a) is captured by the BWI data. Nevertheless, it is seen that the rapid increase in the BWI occurs approximately 1 hour earlier than that in the ICC cumulative heat data. This shift indicates that there are strong chemical reactions occurring even during the induction period (i.e., the initial ~1.5 hours) when the heat flow is low.

To illustrate this point, we plotted the BWI, FWI and CWI against the cumulative heat in Figure 8b, where a continuous increase of BWI and a decrease of FWI are seen with increasing cumulative heat. Both the BWI and FWI are seen to be almost linearly correlated with the ICC cumulative heat during the initial hour (refer to Figure S4 in the Supporting Information for a zoom of the first 2.5 hours data) and after ~2.5 hours, although their slopes of increase/decrease change considerably between the early-age behavior (initial hour) and later stage (~2.5-11.5 hours). The linear correlations in both periods suggest that they each contain a dominant chemical process (denoted by the two dashed lines in Figure 8b for the BWI and FWI), however, the dramatic difference in slope between these periods indicates that the actual processes are dissimilar. Furthermore, the CWI remains essentially unchanged after ~2.5 hours, which suggests that the alkali-activation reaction after ~2.5 hours is characterized by a conversion of FWI to BWI for NaOH-activated slag. This conversion of FWI to BWI after ~2.5 hours is mainly attributed to the formation of the main reaction product in the NaOH-activated slag (i.e., C-(N)-A-S-H gel) where free $H_2O$ is consumed and transformed to chemically bound water in the resulting gel. The formation of a secondary phase (e.g., a hydrotalcite-like phase) may also contribute to this conversion, however, its contribution is likely to be small due to its relatively small quantity of formation after the main reaction peak (i.e., after ~3 hours), as illustrated by a recent quantitative XRD study on NaOH-activated slags.[12]

The increased formation of C-(N)-A-S-H gel after ~2.5 hours in the NaOH-activated slag (Figure 8b) is supported by the FTIR data given in Figure 8c, where the evolution of the main asymmetric Si−O−T stretching band for the same sample over a period of 4 days is displayed. During the initial ~2 hours, no obvious change to the spectrum is observed which indicates that the extent of gel formation is low. However, after ~2 hours, the data show clear emergence and growth of new structural features at ~805, ~895, ~930 and ~985 cm$^{-1}$, that are consistent with the formation of C-S-H-type gels and tobermorite structures.[78-80] The assignment of $Q^n$ silicate species of C-S-H-type gels to specific FTIR bands is far from straightforward, as exhibited in the literature via conflicting assignments,[78, 79, 81] however, based on our data and key literature we have assigned the main band at ~930 cm$^{-1}$, along with the two shoulders at ~895 and ~985 cm$^{-1}$, to asymmetric Si−O stretching in $Q^2$ tetrahedra associated with different neighboring environments.[78, 79] Deconvolution of the FTIR spectra (refer to Figures S5 and S6 in the Supporting Information) shows that the relative intensities of these bands increase/decrease almost linearly with the corresponding BWI/FWI. Hence, these results clearly demonstrate that the linear conversion of FWI to BWI after ~2.5 hours (Figure 8b) is predominately due to precipitation of the C-(N)-A-S-H gel.

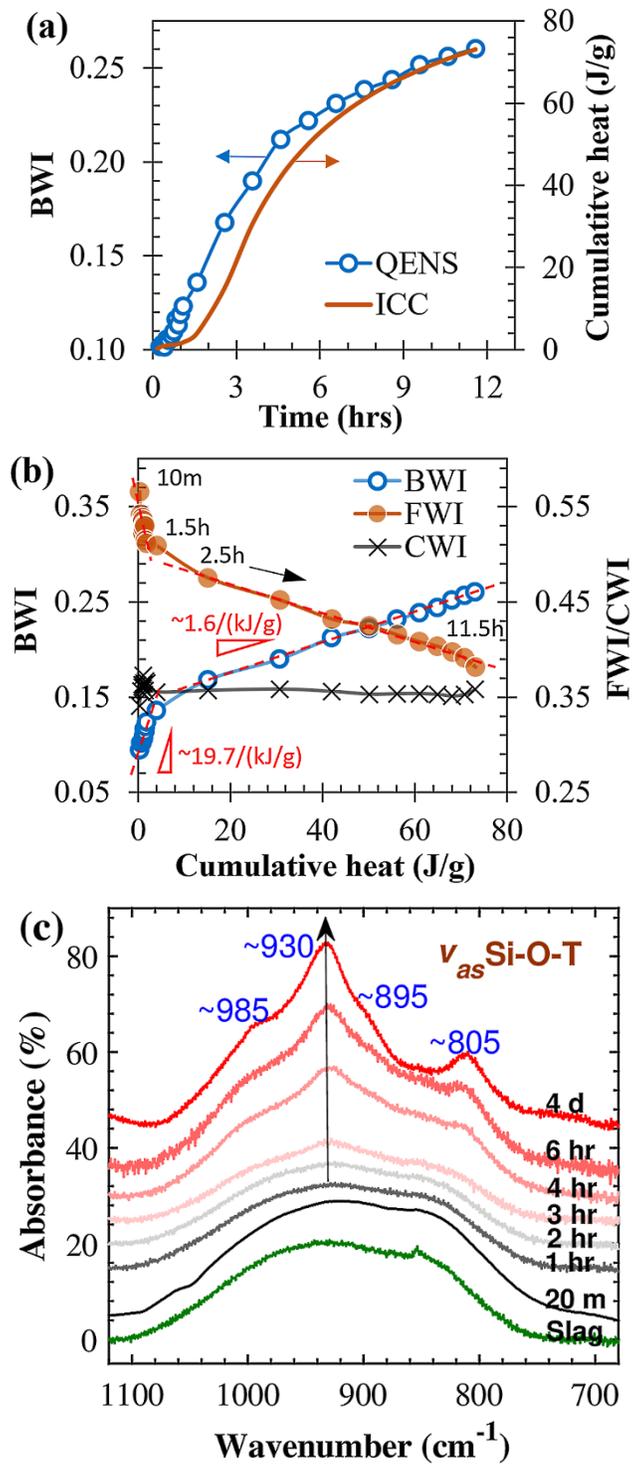

Figure 8. (a) Evolution of bound water index (BWI) and ICC cumulative heat with the progress of reaction, and (b) correlations between ICC cumulative heat and BWI, free water index (FWI) and constrained water index (CWI) for the NaOH-activated slag. (c) FTIR data showing the evolution

of the main asymmetric Si−O−T (T = Si or Al) stretching band for the sample as a function of reaction time up to 4 days. An FTIR spectrum of the raw slag is also shown in (c).

In contrast to the correlation between BWI/FWI and cumulative ICC/FTIR data for NaOH-activated slag after ~2.5 hours, there are limited changes in the FTIR (Figure 8c) and cumulative ICC data (Figures 7 and 8a) during the initial hour of reaction while at the same time the BWI and FWI are dramatically changing (Figure 8b). According to Table 1 and the discussion presented in Section 4.3, one major factor that could lead to the dramatic increase of BWI (as observed in Figure 8b) is an increase in the concentration of silicate species arising from slag dissolution. This is supported by thermodynamic calculations on NaOH-activated slags,[71] where the concentration of silicate species in the pore solution increases dramatically during the initial stage of reaction (due to dissolution of silicate from slag). As silicate is released from the slag (e.g., simplified description of slag silicate dissolution of $SiO_2 + 2OH^- \rightarrow SiO_2(OH)_2^{2-}$) the resulting monomers/oligomers will be protonated to a certain degree according to the pH of the solution and the associated equilibrium constants ($K_a$ values)[82] which will be accompanied by a conversion of FWI to BWI. The increase in silicate species leads to an overall increase of Na+Si concentration in the solution which explains the slight increase of CWI during the initial hour of reaction (Figures 8b and S4 (Supporting Information)).

However, we cannot eliminate the contributions to BWI from two other factors: (1) the formation of C-(N)-A-S-H gel and secondary phases and (2) a potential increase in surface area during the early stages of slag dissolution. Our recent high resolution XRD data (unpublished work) show that a C-(N)-A-S-H-type gel along with secondary hydrotalcite-like and calcium hemicarboaluminate ($Ca_4Al_2(CO_3)_{0.5}(OH)_{13} \cdot 5H_2O$, hereinafter referred as Hc) phases form in a NaOH-activated slag immediately after mixing. Another *in situ* quantitative XRD study shows that the formation of secondary hydrotalcite-like phase is even more pronounced than C-(N)-A-S-H gel formation during the early stages of reaction in NaOH-activated slag.[12] A detailed discussion on the impacts of these factors is presented in the Supporting Information, where we discuss that the formation of hydrotalcite-like phase (as evidenced by FTIR data in Figure S7 in the Supporting Information) may contribute to the BWI to the same extent as the silicate released during slag dissolution. Nevertheless, further research is required to quantify the formation of the hydrotalcite-like phase during the initial hour.

One quantity that can be calculated from the data in Figure 8b is the enthalpy associated with the generation of bound H-atoms for the dominant chemical processes outlined above, based on the slopes of the BWI versus ICC cumulative heat plot. The detailed calculations of this quantity are presented in the Supporting Information. The results are summarized in Table 2, which shows that the enthalpy of generating bound H-atoms is ~1.7 kJ/mol during the initial hour and ~20.8 kJ/mol after ~2.5 hours. We also estimated the enthalpy of silicate release during slag dissolution, C-S-H gel formation and precipitation of hydrotalcite-like and Hc phases based on thermodynamic data,[83] according to the following simplified reactions (refer to the Supporting Information for the calculations):

(1) Slag silicate dissolution: $SiO_2 + 2\ OH^- \rightarrow SiO_2(OH)_2^{2-} + 21.9\ kJ/mol$

(2) C-S-H gel formation: $SiO_2 + CaO + 2\ H_2O \rightarrow CaO \cdot SiO_2 \cdot 2H_2O + 114.2\ kJ/mol$

(3) Hydrotalcite formation: $6MgO + Al_2O_3 + CO_2 + 12H_2O \rightarrow Mg_6Al_2CO_3(OH)_{16} \cdot 4H_2O - 118.2\ kJ/mol$

(4) Hc phase formation: $4CaO + Al_2O_3 + 0.5CO_2 + 11.5\ H_2O \rightarrow Ca_4Al_2(CO_3)_{0.5}(OH)_{16} \cdot 5H_2O + 566.2\ kJ/mol$

The thermochemical calculations show that silicate release during slag dissolution and formation of C-S-H gel and Hc phase are exothermic (reactions (1), (2) and (4)) while hydrotalcite formation (according to reaction (3)) is endothermic. Note that $SiO_2(OH)_2^{2-}$ is used in reaction (1) because it has been shown to be the dominant monomeric species at the pH levels of the alkaline solutions used here (> ~13[14, 71]).[82] The enthalpy of generating bound H-atoms for C-S-H gel formation according to reaction (2) is estimated to be ~28.6 (=114.2/4) kJ/mol, which is close to that estimated from our BWI data after ~2.5 hours in Figure 8b (~20.8 kJ/mol). This is consistent with the QENS and FTIR results (Figures 8 and S6 (Supporting Information), respectively) which clearly show that the dominant reaction process in the NaOH-activated slag after ~2.5 hours is the formation of the C-(N)-A-S-H gel. The enthalpy of generating bound H-atoms during the initial hour (~1.7 kJ/mol) is much lower than that estimated from silicate dissolution according to reaction (1) (~21.9/2=11.0 kJ/mol), suggesting that silicate dissolution is not the only major process leading to the dramatic increase of BWI during the initial hour as seen in Figure 8b. The above calculations indicate that another major process could be the formation of hydrotalcite phase (reaction (3)) which exhibits a negative enthalpy of generating bound H-atoms ($-118.2/24 = \sim -4.9$ kJ/mol) and as a result lowers the overall enthalpy value for the initial hour. The potential large impact of hydrotalcite formation on the enthalpy of generating bound H-atoms is supported by our discussion in the Supporting Information, where we show that it is possible for the formation of hydrotalcite to have a large contribution to the BWI increase during the initial hour in the NaOH-activated slag.

Furthermore, the data and calculations presented here suggest that the formation of calcium-bearing phases (via reactions (2) and (4)) is relatively small during the initial hour. Given that calcium content in the pore solution of NaOH-activated slag is also extremely low during the initial stages of reaction,[71] calcium dissolution from slag is likely to be suppressed initially in this sample when the pH is high and the concentrations of dissolved silicate and aluminate are not high enough to take dissolved $Ca^{2+}$ out of the solution via precipitation. The suppression effect of high pH on calcium dissolution has been reported in alkali-activated metakaolin/portlandite[84] and is also supported by our recent work (unpublished data), where it is shown that increasing alkalinity impede the dissolution of portlandite.

Additional information on the processes occurring during NaOH activation of slag can be obtained by analyzing the evolution of the CWI. Both thermodynamic calculations[71] and pore solution analysis[77] have shown that the Na+Si concentration in the pore solution of NaOH-activated slags decreases continuously during the formation of the main binder gel when both $Na^+$ ions and silicate species are incorporated into the reaction product (e.g., C-(N)-A-S-H gel). This reduction in Na+Si concentration will be accompanied by a reduction in the CWI according to Figure 6b, however, Figure 8b clearly shows that the CWI remains unchanged after ~1.5 hours. According to Table 1 and the discussion presented in Section 4.3, other contributors to the CWI are $H_2O$ absorbed on pore surfaces or confined in small gel pores (~2-5 nm in diameter). The $N_2$ sorption data in Table 3 and Figure S8 in the Supporting Information show that the pores developed after 6-12 hours in this sample are predominantly capillary pores (> 10 nm) with a small proportion of gel pores (< 5 nm), which is consistent with our recent SANS investigation on a NaOH-activated slag paste.[32] As a result, the impact of pore confinement is limited, however, emergence of these pores and the associated increase in the overall pore surface area (see $N_2$ sorption data in Table 3) will lead to an increase in the CWI due to the associated increase in overall percentage of $H_2O$ molecules adsorbed to pore walls. As a result of this increase in adsorbed $H_2O$ (increase in CWI) and decrease in Na+Si concentration (decrease in CWI), the CWI remains essentially unchanged over the period of ~1.5-11.5 hours.

### 4.5.2  $Na_2SiO_3$-activated slag

The evolution of BWI with the progress of reaction for the $Na_2SiO_3$-activated slag is shown in Figure 9a, along with the ICC cumulative heat data. As was the case for the NaOH-activated slag sample, the BWI of the $Na_2SiO_3$-activated slag is seen to follow a similar trend to the cumulative heat curve as the reaction progresses and appears to capture the two reaction peaks seen in the ICC heat flow data (Figure 7). Although the BWI captures the initial rise in the cumulative heat data (i.e., the first reaction peak in heat flow data at ~1 hour), the second rise in the BWI is seen to shift to ~1 hour earlier compared with the ICC data, as also seen for the NaOH-activated slag (Figure 8a). Figure 9b shows the correlations between the three water indices (i.e., BWI, CWI and FWI) and the corresponding ICC cumulative heat for the $Na_2SiO_3$-activated slag sample, where it is clear that the $Na_2SiO_3$-activated slag behavior is distinctly different from that of the NaOH-activated slag sample (Figure 8b). Specifically, the FWI and CWI for $Na_2SiO_3$-activated slag are seen to be inversely correlated as a function of reaction time (and ICC cumulative heat) with local minima/maxima at ~2-3 hours. Furthermore, the data in Figure 9b show that the first heat flow peak in the ICC data (Figure 7) is associated with a conversion of CWI to BWI and FWI. In contrast, after ~2.5 hours, both the BWI and CWI increases as a function of reaction time (and cumulative heat) at the expense of FWI (refer to Figure S9 in the Supporting Information for the evolution of CWI and FWI with time).

#### 4.5.2.1  Before ~2.5 hours

According to the results and discussion presented in Section 4.3 (including Table 1 and Figure 6), the main factor controlling the FWI and CWI during the initial stage when the extent of reaction is low (and hence the specific surface area and existence of small gel pores are also small) is the Na+Si concentration in the pore solution of the AASs. For the $Na_2SiO_3$-activated slag sample presented in Figure 9, the Na+Si concentration of the activating solution is 8.4 M, which is within the region where the CWI is directly proportional to the Na+Si concentration (Figure 6b). Hence, the large reduction in CWI during the initial ~2.5 hours (as shown in Figures 9b and S9 (Supporting Information)) can be attributed to a reduction in the Na+Si concentration in the pore solution. Similarly, the FWI is shown to be inversely related with the Na+Si concentration (Figure 6c), and hence the increase in FWI during the initial ~2-3 hours (as illustrated in Figures 9b and S9 (Supporting Information)) also suggests a reduction in Na+Si concentration in the pore solution over this period. This reduction in Na+Si concentration during the initial stage of reaction for the $Na_2SiO_3$-activated slag is supported by thermodynamic calculations[71] and pore solution analysis,[77] which showed that both the concentration of $Na^+$ ions and silicate species decrease dramatically during the initial stage of reaction for this type of AASs.

There are several possible reaction pathways for the reduction in the concentration of $Na^+$ ions and silicate species in the $Na_2SiO_3$-activated slag sample during the initial stage of reaction. One possibility is polycondensation between silicate species in the original activator solution and Si/Al species dissolved from slag particles, with $Na^+$ incorporated into the condensed gel to charge balance the aluminate species. This polycondensation reaction is similar to what happens when an $Na_2SiO_3$ solution is mixed with metakaolin, however, this reaction tends to reduce the BWI since it converts chemically bound H-atoms in silicate and aluminate monomers and aluminosilicate oligomers to free $H_2O$,[83, 85] as also evidenced by our QENS data on $Na_2SiO_3$-activated metakaolin (unpublished work). This contradicts the BWI results in Figures 9a and 9b, which clearly show a linear increase in the BWI during the initial ~2.5 hours. The initial slope of increase in the BWI as a function of the cumulative heat (and hence the enthalpy of generating bound H-atoms, as shown in Table 2) is closer to the time period of ~2.5 to 11.5 hours for the NaOH-activated slag rather than its initial ~1 hour. This observation indicates that a more probable reaction pathway responsible for the reduction in the pore solution Na+Si concentration of $Na_2SiO_3$-activated slag during the initial ~2.5 hours is the formation of a C-S-H-type gel via reactions between silicate species in the activator solution and $Ca^{2+}$ ions dissolved from slag.

The involvement of $Ca^{2+}$ ions during the first ICC reaction peak of the $Na_2SiO_3$-activated slag (Figure 7b) is supported by thermodynamic calculations[71] and pore solution analysis[77] which showed that $Na_2SiO_3$-activated slag experiences a significant increase in the $OH^-$ concentration during the initial stage of reaction. The OH⁻ increase is mainly attributed to dissolution of calcium (and some magnesium) from slag since dissolution of aluminate and silicate actually consume OH⁻, as illustrated in Table S1 in the Supporting Information. However, the concentration of $Ca^{2+}$ ions in the pore solution is maintained at an extremely low level compared with other ions (3-4 orders of magnitude lower than the concentrations of $Na^+$ ions and silicate species),[71, 77] suggesting that $Ca^{2+}$ ions have been taken out of the solution via product formation, promoting further dissolution of calcium from slag. This reaction route is similar to that of $Na_2CO_3$- and $Na_2SO_4$-activated slags, where the dissolved $Ca^{2+}$ ions from slag particles react with the anions (e.g., $CO_3^{2-}$ and $SO_4^{2-}$) to form calcium carbonate/sulfate precipitates, and the OH⁻ ion concentration increases due to calcium dissolution. This leads to an increase in the solution pH and enhancement of slag dissolution (involving the release of silicate and aluminate species) which enable for the subsequent formation of C-(N)-A-S-H gel (corresponding to the 2$^{nd}$ ICC reaction peak in Figure 7). The ICC data for $Na_2CO_3$- and $Na_2SO_4$-activated slags also exhibit two reaction peaks (similar to the ICC curve of the $Na_2SiO_3$-activated slag in Figure 7), where the first peak has been assigned to precipitation of carbonate and sulfate salts (e.g., calcite and ettringite, respectively).[14, 15, 86-88] Hence, it is clear that the alkali salts/silicates used during alkali activation play an extremely important role in dictating the formation mechanisms in AASs. Moreover, it has been shown that the first reaction peak disappeared when using a solid source of $Na_2SiO_3$ where the silicate in the solid $Na_2SiO_3$ precursor is not readily available upon mixing.[20] Therefore, it is reasonable to attribute the first reaction peak of the $Na_2SiO_3$-activated slag studied here (Figure 7) to the formation of a C-S-H-type gel via reaction between dissolved $Ca^{2+}$ ions and silicate species in the activator solution.

According to thermodynamic modeling,[71] aluminate concentration in the pore solution decreases continuously after the initial increase (due to dissolution), indicating that the aluminate species have been incorporated into reaction products. Due to the abundance of silicate species in the solution, it is likely that the dissolved aluminate species react quickly with silicate species and are incorporated into the gel. In fact, NMR data on silicate-activated slag has showed an increase in $Q^2$(1Al) species after the first ICC reaction peak.[14] $Na^+$ ions are also likely to be incorporated into the gel to charge balance the aluminate sites (since the $Na^+$ concentration in the pore solution is also shown to decrease immediately after mixing[71]), making it a C-(N)-A-S-H-type gel. It is noted that the removal of $Ca^{2+}$ ions and silicate and aluminate species from the pore solution could be partially caused by the formation of secondary phases (e.g., a hydrotalcite-like phase), which have been observed experimentally[9] and predicted to form in silicate-activated slag using thermodynamic calculations.[71]

The formation of a C-(N)-A-S-H-type gel due to the interactions between (i) dissolved $Ca^{2+}$ and aluminate species from slag and (ii) the $Na^+$ and silicate species originally in solution is further supported by the FTIR data in Figure 9c, where the evolution of the main $v_{as}$ Si−O−T band for the $Na_2SiO_3$-activated slag over a period of ~2 days is presented together with the spectra for neat slag and the $Na_2SiO_3$ activating solution. The $Na_2SiO_3$ activating solution exhibits a peak at ~984 cm$^{-1}$ and a shoulder at ~918 cm$^{-1}$, which is consistent with literature data[45, 89] and can be tentatively assigned to $v_{as}$ Si−O−Si in $Q^2$ and $Q^1$, respectively.[90] Both bands are seen to shift continuously to smaller wavenumbers (towards ~965 and ~905 cm$^{-1}$ at ~3 hours) with the progress of reaction during the initial ~3 hours. Similar band shifts to smaller wavenumbers have been reported in the literature for silicate-activated slags and fly ash.[45, 81, 88] It is well established that in silicate gel an increase of Na, Ca or Al content lowers the wavenumber of the $v_{as}$ Si−O−Si band.[21, 78, 89] Therefore, the shifts observed during the initial 2-3 hours in Figure 9c can be attributed to reactions between dissolved $Ca^{2+}$ ions and aluminate species from slag with silicate species and $Na^+$ ions in the $Na_2SiO_3$ solution, leading to formation of a C-(N)-A-S-H-type gel.

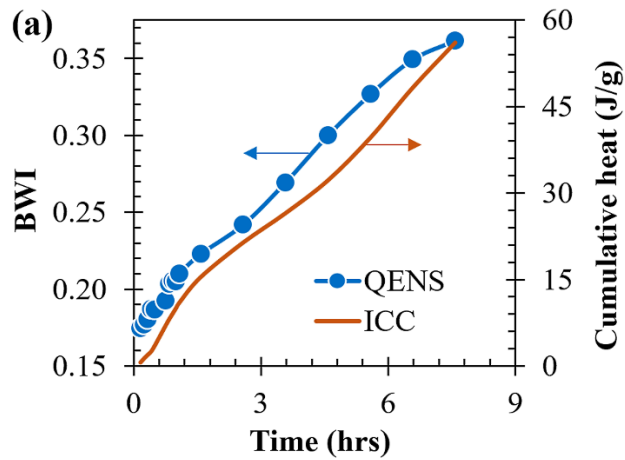

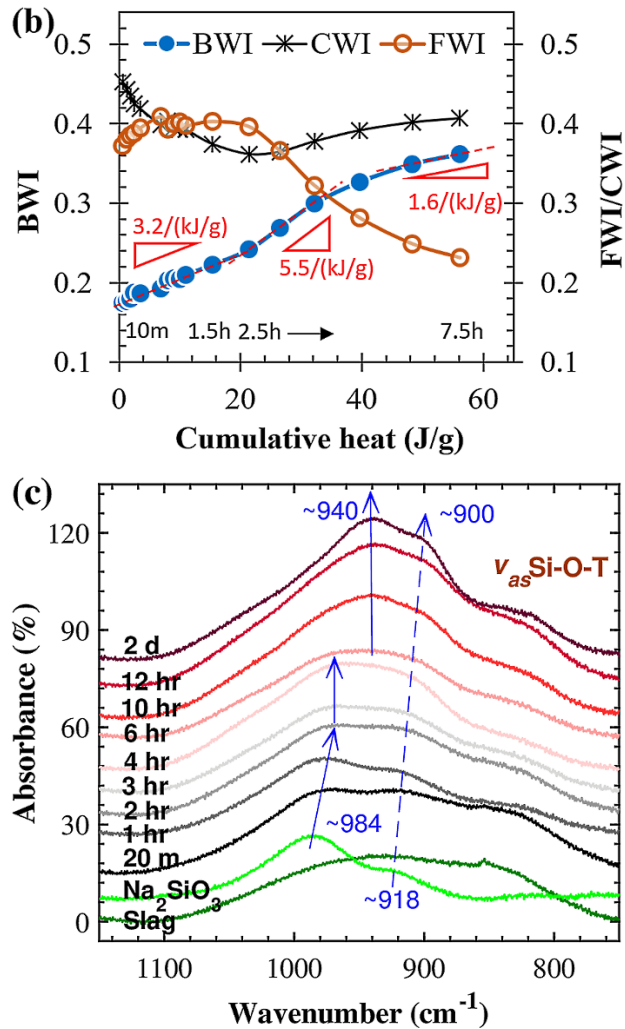

Figure 9 (a) Evolution of bound water index (BWI) and ICC cumulative heat with the progress of reaction and (b) correlations between the BWI, free water index (FWI) and constrained water index (CWI) and the ICC cumulative heat for the Na$_2$SiO$_3$-activated slag sample. (c) FTIR data showing the evolution of the main asymmetric Si−O−T (T = Si or Al) stretching band for the Na$_2$SiO$_3$-activated slag over a period of 2 days. FTIR spectra of the raw slag and the original Na$_2$SiO$_3$ solution are also shown in (c).

### 4.5.2.2 After ~2.5 hours

After ~2.5 hours, the Na$_2$SiO$_3$-activated slag exhibits a continuous transformation of FWI to BWI (Figure 9b), similar to the NaOH-activated slag shown in Figure 8b. The BWI versus ICC cumulative heat plot in Figure 9b clearly shows a region with a steeper slope (~2.5-4.5 hours) before the slope decreases to a level (~1.6 per kJ/g) similar to that of the NaOH-activated slag between ~2.5-11.5 hour. Furthermore, as shown in Table 2, both samples have similar enthalpy values of generating bound H-atoms for this period ( −20.8-21.6 kJ/mol). These results suggest that the Na$_2$SiO$_3$-actived slag is dominated by C-(N)-A-S-H gel precipitation after ~6.5 hours, similar to that of the NaOH-activated slag at ~2.5-11.5 hours. The extensive formation of C-(N)-A-S-H gel after ~6.5 hours is supported by the FTIR results in Figure 9c, which show the emergence of a broad band at ~940 cm$^{-1}$ (attributed to $v_{as}$ Si−O−T stretching in C-(N)-A-S-H gel) at ~6 hours and continuous growth of this band afterwards. The wavenumber of this $v_{as}$ Si−O−T band is slightly higher in the Na$_2$SiO$_3$-activated sample (i.e., ~940 cm$^{-1}$) than in the NaOH-activated sample (i.e., ~930 cm$^{-1}$, Figure 8c). This suggests that the C-(N)-A-S-H gel formed in the former is slightly more polymerized than that form in the latter, which is consistent with NMR data.[77]

Table 2. Summary of the calculated enthalpy of generating bound H-atoms for the different linear periods of the bound water index (BWI) versus ICC cumulative heat plot seen in Figures 8b and 9b. Detailed calculations are shown in the Supporting Information.

| Samples | Time (hours) | Slope of the BWI curves in Figures 8b and 9b (/(kJ/g)) | Enthalpy per mole of bound H-atom generated (kJ/mol) |
|---|---|---|---|
| NaOH-slag | 0-1.0 | 19.7 | 1.7 |
|  | 2.5-11.5 | 1.6 | 20.8 |
| Na$_2$SiO$_3$-slag | 0-2.5 | 3.2 | 10.8 |
|  | 2.5-4.5 | 5.4 | 6.4 |
|  | 6.5-7.5 | 1.6 | 21.6 |

The steeper slope for the BWI during the ~2.5-4.5 hours (~5.4 per kJ/g; Figure 9b) and hence lower value of enthalpy of generating bound H-atoms (~6.4 kJ/mol) can be attributed to an increase of slag dissolution that is driven by the increase in the OH$^-$ ion concentration as already explained above. However, this slope (~5.4 per kJ/g) is still much smaller than that during the initial hour for the NaOH-activated sample in Figure 8b (~19.7 per kJ/g). This can be attributed to the lower OH$^-$ ion concentration in the Na$_2$SiO$_3$-activated slag sample[71] resulting in a slower rate of slag dissolution. Nevertheless, the increased dissolution rate of slag between ~2.5-4.5 hours provides additional silicate species for the formation of C-(N)-A-S-H gel, which leads to the second reaction peak at ~6 hours in the Na$_2$SiO$_3$-activated slag (Figure 7a). Meanwhile, after a period of more rapid slag dissolution that consumes $OH^-$ ions, the rate of dissolution slows down due to (1) the reduction in $OH^-$ ion concentration (as predicted in ref. [71]) and (2) the formation of reaction products on slag surfaces that hinder ion diffusion (it has been shown that the 2nd ICC reaction peak in $Na_2SiO_3$-activated slag is diffusion-controlled[20]).

The $Na_2SiO_3$-activated slag also exhibits an obvious increase in the CWI after ~2.5 hours (Figure 9b), although the Na+Si concentration of the pore solution has been shown to continuously decrease as reaction progresses,[71] which tend to decrease CWI. As explained previously in relation to Table 1, the other factors that could increase CWI are the increase in solid surface area due to formation of reaction products and the emergence of gel pores (~2-5 nm, due to formation of C-(N)-A-S-H gel). Hence, the evident increase in the CWI after ~2.5 hours (Figure 9b) suggests that the latter two factors (i.e., increase in surface area and gel porosity) dominates the $Na_2SiO_3$-activated slag during this stage of the reaction. This is supported by the $N_2$ sorption data in Table 3, which show that the BET surface area of the $Na_2SiO_3$-activated slag is over four times higher than the NaOH-activated slag and increases much faster during ~6-12 hours. Furthermore, the relative amount of gel and capillary pores (Table 3) and the pore size distributions (Figure S8 in Supporting Information) derived from the BJH adsorption branch show that around 9% of the pore volume accessible using $N_2$ sorption in the 6-hour $Na_2SiO_3$-activated slag is from gel pores (d < 5 nm). This indicates that a fair amount of C-(N)-A-S-H-type gel has already formed before the main ICC reaction peak at ~ 6 hour (Figure 7a), in agreement with the QENS and FTIR data. At 12 hours, the relative amount of pores accessible using $N_2$ sorption with diameters smaller than 5 nm increases to over 50% (Table 3 and Figure S8 in the Supporting Information). The development of gel pores in the $Na_2SiO_3$-activated slag (as opposed to capillary pores for the NaOH-activated slag) is consistent with recent SANS measurements on similar AASs.[32] Therefore, the increase in CWI for the $Na_2SiO_3$-activated slag after ~2.5 hours (Figure 9b) is attributed to a significant increase in pore surface area and gel porosity that suppresses the inverse impact from a decrease of the Na+Si concentration.

Table 3. Summary of the BET surface area and the relative amount of gel and capillary pores (based on the BJH adsorption branch) from $N_2$ sorption measurements of the NaOH- and $Na_2SiO_3$-activated slag samples cured for ~6 and ~12 hours.

| Sample | Age (hours) | BET surface area (m$^2$/g) | BJH gel pores d < 5 nm | BJH capillary pores d > 10 nm |
|---|---|---|---|---|
| NaOH-slag | 6 | 14.0 | 3% | 92%* |
| NaOH-slag | 12 | 15.7 | 5% | 81%* |
| Na$_2$SiO$_3$-slag | 6 | 54.1 | 9% | 50% |
| Na$_2$SiO$_3$-slag | 12 | 109.2 | 53% | 6% |

* These values could be even higher since NaOH-activated slag sample contains pores larger than what is accessible using N$_2$ sorption (d < 50 nm) as shown in Figure S8a in the Supporting Information.

### 4.6 Broader Implications

Based on the detailed analysis outlined above of the QENS-INS, FTIR and N$_2$ sorption results for Na$_2$SiO$_3$-activated slag (Figure 9), together with the thermodynamic data in the literature on similar AASs,[71] it is clear that both the first and second ICC reaction peaks in this sample (as shown in Figure 7) are attributed to the formation of C-(N)-A-S-H-type gels. Nevertheless, it is unclear based on the current study whether the initial gel formed during the first reaction peak possesses the same attributes (e.g., Ca/Si, degree of silica polymerization and incorporation of Na and alumina species) as the main gel formed at a later stage. The formation of the first gel in the initially solution occupied space is associated with several distinct properties of Na$_2$SiO$_3$-activated slag that are different from a NaOH-activated slag; for example, faster initial and final setting,[14, 88] finer pore structure,[32, 46] greater long-term strength,[75, 88] significantly lower intrinsic permeability,[46] and larger extent of drying shrinkage.[91] The formation mechanism of the first gel possesses key similarities to the mechanisms occurring in Na$_2$CO$_3$- and Na$_2$SO$_4$-activated slags, where the dissolution of Ca$^{2+}$ from slag is promoted via precipitation of calcium carbonate/sulfate phases.[14, 19] This mechanism is intrinsically different from NaOH-activated slag where calcium dissolution is hindered initially due to the high pH[84] and the absence of abundant anions in the original solutions (e.g., SiO$_2$(OH)$_2^{2-}$, CO$_3^{2-}$, and SO$_4^{2-}$) to enhance the release of Ca$^{2+}$.

These moderate activators (e.g., Na$_2$SO$_4$ and Na$_2$CO$_3$) can be obtained directly from mining, and hence they are less expensive to produce and have significantly lower carbon footprints compared with silicate and hydroxide activating solutions.[4, 5] However, AASs based on these moderate activators are often associated with delayed setting time and strength development,[14, 19, 87, 88] owing to their lower alkalinity (hence slower dissolution of silicate and aluminate from slag) and the less cohesive nature of the calcium carbonate/sulfate phases that form during the first ICC reaction peak (as opposed to C-(N)-A-S-H-type gel for silicate activating solution). The evident impact of the different activating solutions on the fresh and hardened properties of AASs offers enormous opportunities to tailor the chemistry of the activating solutions (e.g., mixing of different solutions,[20, 88] and use of additives (e.g., nanoparticles)[39, 86, 92]) to achieve desired properties and performance for specific applications, while at the same time lowering the cost and carbon footprint of the resulting AASs. To fully exploit these benefits, it is necessary to elucidate the reaction mechanisms occurring in these complex AAS mixtures, and the current study present an excellent example of achieving this goal by combining *in situ* QENS-INS with laboratory-based experimental techniques.

In particular, the unique combination of *in situ* QENS with ICC data allows for the enthalpy of reaction (in kJ/mol of bound H-atoms) for each reaction stage to be determined. This novel methodology could be readily transferred to the study of other important reaction processes that are accompanied with changes of $H/H_2O$ dynamics (e.g., converting free $H_2O$ to bound $H/H_2O$ and vice versa) as reaction proceeds. It is also important to note that broadband neutron spectrometers, such as VISION (~ −2 to ~1000 meV), are ideal for studying $H/H_2O$ dynamics in systems where extremely high energy resolution of the QENS component is not required (i.e., systems containing a wide range of H-atom environments; specifically, free, constrained and bound water). Although VISION is not a QENS-specific spectrometer, this study has shown that the QENS component is readily accessible and analysis of the QENS data provides important mechanistic insight on the alkali-activation reaction that complements the water libration mode contained within the INS component.

## 5  Conclusions

In this study, *in situ* quasi-elastic neutron scattering (QENS) technique has been employed to probe the water dynamics and formation mechanisms of alkali-activated slags (AASs). A double-Lorentzian model was used to fit the QENS spectra, from which three water indices (i.e., bound water index (BWI), constrained water index (CWI) and free water index (FWI)) were derived. Comparison of the QENS water indices with isothermal conduction calorimetry (ICC) data showed that the evolution of BWI captures both the single and the double ICC reaction peaks in the NaOH- and Na$_2$SiO$_3$-activated slag, respectively. Analysis of the different water indices together with the ICC and Fourier transform infrared spectroscopy (FTIR) data for the NaOH-activated slag revealed the existence of two distinct stages during the formation process where different reaction processes may have dominated: (1) a faster transformation of FWI to BWI during the initial hour that is mainly attributed to silicate release during slag dissolution and the formation of a hydrotalcite-like phase, and (2) a slow and steady transformation of FWI to BWI after ~2.5 hours arising from precipitation of C-(N)-A-S-H gel.

In contrast, for the Na$_2$SiO$_3$-activated slag sample, the evolution of water indices reveals three reaction stages: an initial stage (up to ~2.5 hours) characterized by a conversion of CWI to BWI and FWI, and a second (~2.5-4.5 hours) and third (after ~6.5 hours) stage both involving a continuous transformation of FWI to BWI and CWI. Analysis of the QENS, FTIR and N$_2$ sorption data along with thermodynamic data in the literature revealed that the first ICC reaction peak (at ~40 min) is mainly attributed to the formation of a C-(N)-A-S-H-type gel, resulting from reactions between dissolved Ca$^{2+}$/aluminate from slag and Na$^+$/silicate in the original Na$_2$SiO$_3$ solution. The third stage of reaction (corresponds to the second ICC reaction peak at ~6 hours) is also attributed to the formation of a C-(N)-A-S-H-type gel, however, the analysis indicates that the main source of silicate for the formation of this gel is from slag dissolution, which is seen to dominate the formation mechanism during the second stage of reaction (~2.5-4.5 hours).

In summary, the current study has demonstrated that *in situ* QENS is a valuable technique for elucidating the detailed formation mechanisms of different types of alkali-activated slags. For the first time, strong experimental evidence has been provided for the formation of an initial C-(N)-A-S-H-type gel in the Na$_2$SiO$_3$-activated slag. Furthermore, the unique combination of QENS and ICC data enabled for the estimation of the enthalpy of reaction (in kJ/mol of bound H-atoms generated), an approach that will be of significant interest to other materials systems involving important reaction processes where the H-atom dynamics change with the progress of reaction.

# 6 Supporting Information

Fits of the quasi-elastic neutron scattering data with a single-Lorentzian model; Inelastic neutron scattering data: Water librational peak; Estimation of the average number of $H_2O$ molecules solvating each ion in alkaline solutions; Isothermal conduction calorimetry (ICC) data for a metakaolin-water mixture; BWI/CWI/FWI versus cumulative heat for NaOH-activated slag – Initial 2.5 hours; Analysis of the FTIR data for NaOH-activated slag; Surface area and hydrotalcite formation during the initial hour of reaction in the NaOH-activated slag and their potential impacts on BWI; Determination of the enthalpy of generating bound H-atoms; BJH pore size distribution from $N_2$ sorption; Evolution of CWI and FWI as a function of time for the $Na_2SiO_3$-activated slag.

# 7 Acknowledgments

This work was supported by the National Science Foundation under Grant No. 1362039. KG's participation was enabled by a Charlotte Elizabeth Procter Fellowship from Princeton University. The authors would like to acknowledge the use of the VISION beamline at the Spallation Neutron Source, a DOE Office of Science User Facility operated by the Oak Ridge National Laboratory. The authors would like to thank Mr. Eric R. McCaslin, Dr. Timmy Ramirez-Cuesta and Dr. Eugene Mamontov for interesting discussions and/or suggestions on data analysis.

# 8 Conflicts of Interest

There are no conflicts to declare.

# 9 References

1. M. Komljenović, in *Handbook of alkali-activated cements, mortars, and concretes*, eds. F. Pacheco-Torgal, J. A. Labrincha, C. Leonelli, A. Palomo and P. Chindaprasirt, Woodhead Publishing, Cambridge, UK, 2015, pp. 171-217.
2. J. A. Ober, *Mineral commodity summaries 2018*, US Geological Survey, Reston, Virginia, USA, 2018.
3. P. J. M. Monteiro, S. A. Miller and A. Horvath, *Nat. Mater.*, 2017, **16**, 698-699.
4. G. Habert and C. Ouellet-Plamondon, *RILEM Tech. Lett.*, 2016, **1**, 17-23.
5. J. L. Provis and S. A. Bernal, *Annu. Rev. Mater. Res.*, 2014, **44**, 299-327.
6. J. L. Provis and J. S. J. van Deventer, eds., *Alkali activated materials: state-of-the-art report, RILEM TC 224-AAM*, Springer/RILEM, Dordrecht, 2014.
7. J. L. Provis and J. S. J. Van Deventer, eds., *Geopolymers: Structures, processing, properties and industrial applications*, Woodhead Publishing Limited, Cambridge, UK, 2009.
8. K. Gong and C. E. White, *J. Phys. Chem. C*, 2018, **122**, 5992–6004.

9. S. A. Bernal, R. San Nicolas, R. J. Myers, R. M. de Gutiérrez, F. Puertas, J. S. J. van Deventer and J. L. Provis, *Cem. Concr. Res.*, 2014, **57**, 33-43.
10. Z. Zhang, J. L. Provis, H. Wang, F. Bullen and A. Reid, *Thermochim. Acta*, 2013, **565**, 163-171.
11. Z. Zhang, H. Wang, J. L. Provis, F. Bullen, A. Reid and Y. Zhu, *Thermochim. Acta*, 2012, **539**, 23-33.
12. Z. Sun and A. Vollpracht, *Cem. Concr. Res.*, 2018, **103**, 110-122.
13. D. Krizan and B. Zivanovic, *Cem. Concr. Res.*, 2002, **32**, 1181-1188.
14. A. Fernández-Jiménez and F. Puertas, *Adv. Cem. Res.*, 2001, **13**, 115-121.
15. C. Shi and R. L. Day, *Cem. Concr. Res.*, 1995, **25**, 1333-1346.
16. K. Yang and C. E. White, *Langmuir*, 2016, **32**, 11580-11590.
17. C. E. White, J. L. Provis, B. Bloomer, N. J. Henson and K. Page, *Phys. Chem. Chem. Phys.*, 2013, **15**, 8573-8582.
18. C. E. White, K. Page, N. J. Henson and J. L. Provis, *Appl. Clay Sci.*, 2013, **73**, 17-25.
19. S. A. Bernal, J. L. Provis, R. J. Myers, R. San Nicolas and J. S. van Deventer, *Mater. Struc.*, 2015, **48**, 517-529.
20. D. Ravikumar and N. Neithalath, *Thermochim. Acta*, 2012, **546**, 32-43.
21. A. Hajimohammadi, J. L. Provis and J. S. J. Van Deventer, *J. Colloid Interf. Sci.*, 2011, **357**, 384-392.
22. J. W. Bullard, H. M. Jennings, R. A. Livingston, A. Nonat, G. W. Scherer, J. S. Schweitzer, K. L. Scrivener and J. J. Thomas, *Cem. Concr. Res.*, 2011, **41**, 1208-1223.
23. S. A. FitzGerald, D. A. Neumann, J. J. Rush, D. P. Bentz and R. A. Livingston, *Chem. Mater.*, 1998, **10**, 397-402.
24. J. J. Thomas, S. A. FitzGerald, D. A. Neumann and R. A. Livingston, *J. Am. Ceram. Soc.*, 2001, **84**, 1811-1816.
25. FitzGerald S.A., Thomas J.J. and Neumann D.A., *Cem. Concr. Res.*, 2002, **32**, 409-413.
26. A. J. Allen, J. C. McLaughlin, D. A. Neumann and R. A. Livingston, *J. Mater. Res.*, 2004, **19**, 3242-3254.
27. V. K. Peterson, D. A. Neumann and R. A. Livingston, *J. Phys. Chem. B*, 2005, **109**, 14449-14453.
28. V. K. Peterson, D. A. Neumann and R. A. Livingston, *Physica B*, 2006, **385-386**, 481-486.
29. V. K. Peterson, in *Studying kinetics with neutrons*, eds. G. Eckold, H. Schober and S. E. Nagler, Springer, Heidelberg, Germany, 2010, pp. 19-75.
30. T. Gutberlet, H. Hilbig, R. E. Beddoe and W. Lohstroh, *Cem. Con. Res.*, 2013, **51**, 104-108.
31. K. Kupwade-Patil, M. Tyagi, C. M. Brown and O. Büyüköztürk, *Cem. Con. Res.*, 2016, **86**, 55-62.
32. C. E. White, D. P. Olds, M. Hartl, R. P. Hjelm and K. Page, *J. Appl. Cryst.*, 2017, **50**, 61-75.
33. K. Kupwade-Patil, S. O. Diallo, D. Z. Hossain, M. R. Islam and E. N. Allouche, *Constr. Build. Mater.*, 2016, **120**, 181-188.
34. C. E. White, J. L. Provis, A. Llobet, T. Proffen and J. S. J. van Deventer, *J. Am. Ceram. Soc.*, 2011, **94**, 3532-3539.
35. J. J. Thomas, A. J. Allen and H. M. Jennings, *J. Phys. Chem. C*, 2009, **113**, 19836-19844.
36. F. Ridi, L. Dei, E. Fratini, S.-H. Chen and P. Baglioni, *J. Phys. Chem. B*, 2003, **107**, 1056-1061.

37. V. K. Peterson and M. C. G. Juenger, *Chem. Mater.*, 2006, **18**, 5798-5804.
38. C. Hesse, F. Goetz-Neunhoeffer and J. Neubauer, *Cem. Concr. Res.*, 2011, **41**, 123-128.
39. N. Garg and C. E. White, *J. Mater. Chem. A*, 2017, **5**, 11794-11804.
40. A. E. Morandeau and C. E. White, *J. Mater. Chem. A*, 2015, **3**, 8597-8605.
41. Q. Hu, M. Aboustait, T. Kim, M. T. Ley, J. W. Bullard, G. W. Scherer, J. C. Hanan, V. Rose, R. Winarski and J. Gelb, *Cem. Concr. Res.*, 2016, **89**, 14-26.
42. J. Skibsted and C. Hall, *Cem. Concr. Res.*, 2008, **38**, 205-225.
43. R. Holly, H. Peemoeller, M. Zhang, E. Reardon and C. Hansson, *J. Am. Ceram. Soc.*, 2006, **89**, 1022-1027.
44. S. Del Buffa, E. Fratini, F. Ridi, A. Faraone and P. Baglioni, *J. Phys. Chem. C*, 2016, **120**, 7612-7620.
45. C. A. Rees, J. L. Provis, G. C. Lukey and J. S. J. Van Deventer, *Langmuir*, 2007, **23**, 9076-9082.
46. A. Blyth, C. A. Eiben, G. W. Scherer and C. E. White, *J. Am. Ceram. Soc.*, 2017, **100**, 1-12.
47. N. W. Ockwig, R. T. Cygan, M. A. Hartl, L. L. Daemen and T. M. Nenoff, *J. Phys. Chem. C*, 2008, **112**, 13629-13634.
48. K. Gong and C. E. White, *Cem. Concr. Res.*, 2016, **89**, 310-319
49. J. W. Phair, R. A. Livingston, C. M. Brown and A. J. Benesi, *Chem. Mater.*, 2004, **16**, 5042-5050.
50. C. Corsaro, V. Crupi, D. Majolino, S. F. Parker, V. Venuti and U. Wanderlingh, *J. Phys. Chem. A*, 2006, **110**, 1190-1195.
51. J. Li, *J. Chem. Phys.*, 1996, **105**, 6733-6755.
52. D. Argyris, N. R. Tummala, A. Striolo and D. R. Cole, *J. Phys. Chem. C*, 2008, **112**, 13587-13599.
53. J. Teixeira, M. C. Bellissent-Funel, S. H. Chen and A. J. Dianoux, *Phys. Rev. A*, 1985, **31**, 1913-1917.
54. S. Mitra, R. Mukhopadhyay, I. Tsukushi and S. Ikeda, *J. Phys. - Condens. Mat.*, 2001, **13**, 8455-8465.
55. H. N. Bordallo, L. P. Aldridge and A. Desmedt, *J. Phys. Chem. B*, 2006, **110**, 17966-17976.
56. T. Takamuku, M. Yamagami, H. Wakita, Y. Masuda and T. Yamaguchi, *J. Phys. Chem. B*, 1997, **101**, 5730-5739.
57. R. W. Impey, P. A. Madden and I. R. McDonald, *J. Phys. Chem.*, 1983, **87**, 5071-5083.
58. N. A. Hewish, J. E. Enderby and W. S. Howells, *J. Phys. C Solid State*, 1983, **16**, 1777.
59. I. C. Bourg and C. I. Steefel, *J. Phys. Chem. C*, 2012, **116**, 11556-11564.
60. D. Presti, A. Pedone, G. Mancini, C. Duce, M. R. Tiné and V. Barone, *Phys. Chem. Chem. Phys.*, 2016, **18**, 2164-2174.
61. D. Hou, Z. Li, T. Zhao and P. Zhang, *Phys. Chem. Chem. Phys.*, 2015, **17**, 1411-1423.
62. E. Chiavazzo, M. Fasano, P. Asinari and P. Decuzzi, *Nat. Commun.*, 2014, **5**.
63. R. T. Cygan, L. L. Daemen, A. Ilgen, J. L. Krumhansl and T. M. Nenoff, *J. Phys. Chem. C*, 2015, **119**, 28005–28019.
64. I. M. Briman, D. Rebiscoul, O. Diat, J.-M. Zanotti, P. Jollivet, P. Barboux and S. Gin, *J. Phys. Chem. C*, 2012, **116**, 7021-7028.
65. S. H. Lee and P. J. Rossky, *J. Chem. Phys.*, 1994, **100**, 3334-3345.
66. J.-P. Korb, P. J. McDonald, L. Monteilhet, A. G. Kalinichev and R. J. Kirkpatrick, *Cem. Concr. Res.*, 2007, **37**, 348-350.

67. Y. Marcus, *Chem. Rev.*, 2009, **109**, 1346-1370.
68. E. Guàrdia, D. Laria and J. Martí, *J. Phys. Chem. B*, 2006, **110**, 6332-6338.
69. S. H. Lee and J. C. Rasaiah, *J. Phys. Chem.*, 1996, **100**, 1420-1425.
70. B. L. Mooney, L. R. Corrales and A. E. Clark, *J. Phys. Chem. B*, 2012, **116**, 4263-4275.
71. B. Lothenbach and A. Gruskovnjak, *Adv. Cem. Res.*, 2007, **19**, 81-92.
72. E. K. Bahlmann, R. K. Harris, K. Metcalfe, J. W. Rockliffe and E. G. Smith, *J. Chem. Soc. Faraday T.*, 1997, **93**, 93-98.
73. R. Mills, *J. Phys. Chem.*, 1973, **77**, 685-688.
74. J. Ramsay, in *The structure, dynamics and equilibrium properties of colloidal systems*, eds. D. M. Bloor and E. Wyn-Jones, Springer, Dordrecht, Netherlands, 1990, pp. 635-651.
75. B. S. Gebregziabiher, R. Thomas and S. Peethamparan, *Cem. Concr. Comp.*, 2015, **55**, 91-102.
76. K. C. Newlands and D. E. Macphee, *Adv. Appl. Ceram.*, 2017, **116**, 216-224.
77. F. Puertas, A. Fernández-Jiménez and M. T. Blanco-Varela, *Cem. Concr. Res.*, 2004, **34**, 139-148.
78. P. Yu, R. J. Kirkpatrick, B. Poe, P. F. McMillan and X. Cong, *J. Am. Ceram. Soc.*, 1999, **82**, 742-748.
79. A. Vidmer, G. Sclauzero and A. Pasquarello, *Cem. Concr. Res.*, 2014, **60**, 11-23.
80. N. V. Chukanov, in *Infrared spectra of mineral species*, Springer, Dordrecht, Netherlands, 2014, pp. 21-1701.
81. A. Dakhane, S. B. Madavarapu, R. Marzke and N. Neithalath, *Appl. Spectrosc.*, 2017, **7**, 1795-1807.
82. J. Šefčík and A. V. McCormick, *AIChE J.*, 1997, **43**, 2773-2784.
83. C. E. White, J. L. Provis, G. J. Kearley, D. P. Riley and J. S. J. van Deventer, *Dalton T.*, 2011, **40**, 1348-1355.
84. S. Alonso and A. Palomo, *Cem. Concr. Res.*, 2001, **31**, 25-30.
85. P. Duxson, A. Fernández-Jiménez, J. L. Provis, G. C. Lukey, A. Palomo and J. S. J. van Deventer, *J. Mater. Sci.*, 2007, **42**, 2917-2933.
86. X. Ke, S. A. Bernal and J. L. Provis, *Cem. Concr. Res.*, 2016, **81**, 24-37.
87. N. Mobasher, S. A. Bernal and J. L. Provis, *J. Nucl. Mater.*, 2016, **468**, 97-104.
88. A. Fernández-Jiménez and F. Puertas, *Adv. Cem. Res.*, 2003, **15**, 129-136.
89. L. Vidal, E. Joussein, M. Colas, J. Cornette, J. Sanz, I. Sobrados, J. L. Gelet, J. Absi and S. Rossignol, *Colloid Surface A*, 2016, **503**, 101-109.
90. A. Gharzouni, E. Joussein, B. Samet, S. Baklouti, S. Pronier, I. Sobrados, J. Sanz and S. Rossignol, *J. Sol-Gel Sci. Techn.*, 2015, **73**, 250-259.
91. C. D. Atiş, C. Bilim, Ö. Çelik and O. Karahan, *Constr. Build. Mater.*, 2009, **23**, 548-555.
92. K. Yang, V. O. Özçelik, N. Garg, K. Gong and C. E. White, *Phys. Chem. Chem. Phys.*, 2018, **20**, 8593-8606.